\documentclass[sigconf]{acmart}
\AtBeginDocument{%
  \providecommand\BibTeX{{%
    \normalfont B\kern-0.5em{\scshape i\kern-0.25em b}\kern-0.8em\TeX}}}

\copyrightyear{2024} 
\acmYear{2024} 
\setcopyright{rightsretained} 
\acmConference[CHI EA '24]{Extended Abstracts of the CHI Conference on Human Factors in Computing Systems}{May 11--16, 2024}{Honolulu, HI, USA}
\acmBooktitle{Extended Abstracts of the CHI Conference on Human Factors in Computing Systems (CHI EA '24), May 11--16, 2024, Honolulu, HI, USA}
\acmDOI{10.1145/3613905.3650989}
\acmISBN{979-8-4007-0331-7/24/05}

\usepackage{wrapfig}
\usepackage{graphicx}
\usepackage{hyperref}
\usepackage{siunitx}
\usepackage{subcaption}

\begin{document}

\title[Has the Virtualization of the Face Changed Facial Perception?]{Has the Virtualization of the Face Changed Facial Perception? A Study
of the Impact of Photo Editing and Augmented Reality on Facial Perception}

\author{Louisa Conwill}
\email{lconwill@nd.edu}
\orcid{0009-0001-7116-266X}
\affiliation{%
  \institution{University of Notre Dame}
  \city{Notre Dame}
  \state{Indiana}
  \country{USA}
}

\author{Sam English Anthony}
\email{sam.e.anthony@gmail.com}
\affiliation{%
  \institution{Independent Researcher}
  \city{Somerville}
  \state{Massachusetts}
\country{USA}
}

\author{Walter J. Scheirer}
\email{wscheire@nd.edu}
\affiliation{%
  \institution{University of Notre Dame}
  \city{Notre Dame}
  \state{Indiana}
  \country{USA}
}

\renewcommand{\shortauthors}{Conwill et al.}

\begin{abstract}
Augmented reality and other photo editing filters are popular methods used to modify faces online. Considering the important role of facial perception in communication, how do we perceive this increasing number of modified faces? In this paper we present the results of six surveys that measure familiarity with different styles of facial filters, perceived strangeness of faces edited with different filters, and ability to discern whether images are filtered. Our results demonstrate that faces modified with more traditional face filters are perceived similarly to unmodified faces, and faces filtered with augmented reality filters are perceived differently from unmodified faces. We discuss possible explanations for these results, including a societal adjustment to traditional photo editing techniques or the inherent differences in the different types of filters. We conclude with a discussion of how to build online spaces more responsibly based on our results.
\end{abstract}




\maketitle

\begin{figure}
  \includegraphics[width=0.4\textwidth]{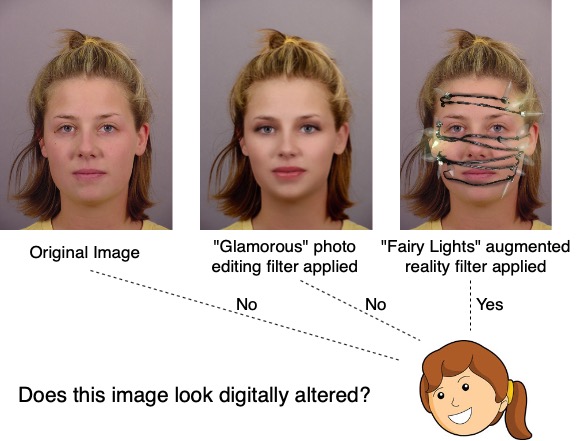}  
  \caption{In our study, we sought to better understand human facial perception of filtered faces. We asked participants to evaluate images of faces with different facial filters applied. In different surveys, we asked participants (1) if the styles of these images are familiar, (2) if the images look strange, and (3) if the images look digitally altered (the question depicted above).}
  \Description{Three images of the same woman. The first image is unfiltered, the second image has a makeup filter on her face, and the final image has an augmented reality filter of Christmas lights overlayed to her face. The images are labeled, ``Original Image,'' ``Glamorous photo editing filter applied,'' and ``Fairy Lights augmented reality filter applied,'' respectively. Below the three images is the following question: ``Does this image look digitally altered?'' A cartoon face is shown with a line from the face to each of the three images. Along the line to the original image and the image with a makeup filter it says, ``No,'' indicating that a possible response to the question could be that neither of these images look digitally altered. Along the line to the image with the lights filter it says, ``Yes,'' indicating that a possible response to the question is that it looks digitally altered.}
  \vspace{-15pt}
  \label{fig:teaser}
\end{figure}

\section{INTRODUCTION}
Facial perception is one of the most sophisticated human visual skills~\cite{haxby2000distributed}. In addition to identifying individuals, the brain's facial recognition abilities allow us to deduce ``a wealth of information that facilitates social communication''~\cite{haxby2000distributed}.  One's particular facial traits affect how they are perceived ~\cite{oosterhof2008functional}. The judgments made from facial appearance, even if inaccurate, have powerful social impacts ranging from electoral success~\cite{todorov2005inferences}~\cite{ballew2007predicting}~\cite{little2007facial} to sentencing decisions~\cite{blair2004influence}~\cite{eberhardt2006looking}.

Computer vision technologies allow us to modify the way our faces look online. As Adobe Photoshop and similar photo-editing tools have evolved over the past thirty years, computer users have been able to edit their faces by changing the image colors, smoothing their skin or whitening their teeth, adding makeup, and even adjusting the size and shape of various facial features. In contrast to these traditional photo editing techniques, more recently, Augmented Reality (AR) filters that overlay virtual content onto the image stream of a smartphone camera have emerged. These filters might add virtual sunglasses to a face or transform one's likeness into that of a dog. Compared to traditional photo editing techniques that quietly modify reality to make it appear more beautiful but still plausible, many (but not all) AR filters intentionally change the image in unrealistic ways, transporting us to new worlds. In 2015, Snapchat was the first social media application to incorporate AR filters into their platform~\cite{inde_2020}. Now, an increasing number of social media platforms including Instagram and TikTok offer AR filters to their users. Filtered photos (using both traditional and AR techniques) now drive a significant amount of engagement on social media~\cite{lavrence2020look}. One recent survey of 18-30 year old young women in the UK found that 90\% of respondents had used such filters~\cite{gill2021changing}. AR beautification filters are becoming so common that both the video conferencing application Zoom and the Chinese mobile payment application Alipay allow users to view and present their faces through beauty filters~\cite{Shein_2021}~\cite{pitcher_2020}~\cite{peng2020alipay}.

Image modification is not without consequence. In the early 2000s, before social media was as ubiquitous as it is today, concern was raised over the prevalence of photoshopped images in magazines~\cite{Wilson_2009}. Now, no longer limited to celebrity images, the trend of sharing modified selfies on social media has garnered media attention for its association with lower self-esteem and body dysmorphia~\cite{Shein_2021}~\cite{fastoso2021mirror}, sometimes referred to as \textit{Snapchat Dysmorphia}~\cite{eshiet2020real} for its association with AR beautification filters on Snapchat. In extreme cases, body dysmorphia from beauty filter usage has pushed people towards plastic surgery~\cite{hunt_2019}. However, some people experience positive effects from sharing filtered images. These include overcoming shyness, mood improvement, and more free self-expression~\cite{pitcher_2020}~\cite{ryan-mosley_2021}. As the tech industry moves towards metaverse experiences, AR usage could become increasingly widespread. A better understanding of our perception of augmented images will help us both better care for the mental health of current AR users and build the metaverse responsibly. 

How we perceive filtered faces may also change over time as we become accustomed to the filters. There is evidence that ``exposure to systematically distorted faces produces corresponding changes in what looks average'' and that the shift in what an average face looks like influences our judgment of facial attractiveness~\cite{rhodes2003fitting}. Because of the influence facial perception has in social communication, there may be social impacts if our exposure to filtered faces online is similarly changing which faces look normal to us.

This paper presents a first step towards an understanding of how filtered faces on the Internet are perceived. We approach this task with three research questions. Our key question is:\\ \textbf{RQ1:} Do filtered faces look normal or strange to people? Does this vary by which filter is used?

Responses to this question could be influenced by familiarity with that style of image filter and ability to tell that a filter has been applied. Thus, we additionally consider the following two questions:\\
\textbf{RQ2:} Which styles of facial filters look familiar to people?\\
    \textbf{RQ3:} Can people tell that filtered images are digitally altered in the first place? 

In this collaboration between computer scientists and psychologists, we present the results of six surveys: one two-alternative forced choice (2AFC) survey and one Likert scale survey for each of the three research questions above. Our results provide insight into how faces filtered with different styles of filters are being perceived. From these insights, we discuss how to design facial filters in socially responsible ways. 

\section{RELATED WORK}

There is a wide body of literature on how photo filtering affects the perception of self and others. The way one presents oneself online is strongly linked to one's self-identity~\cite{jin2012virtual}. AR beautification has the potential to increase self-esteem~\cite{javornik2017mirror}, though in some cases, AR may only change the perception of self when the user is dissatisfied with their appearance~\cite{Yim_Park_2019}. High self-esteem individuals may even resist this \emph{augmented self} because it increases the gap between their ideal attractiveness and actual attractiveness, while low self-esteem individuals may welcome the augmented self~\cite{javornik2021augmented}. Beauty filter users may identify with the filtered version of themselves more than with the real-life version of themselves~\cite{ryan-mosley_2021}.

One study found that people had trouble discriminating between an unmodified picture of their face and a picture of their face edited to have smaller features. Additionally, around a quarter of the study participants preferred pictures of themselves with larger eyes and around half of the participants preferred pictures of themselves with smaller noses~\cite{felisberti2014self}. Taken together this could indicate that filter users may have trouble distinguishing real photos of themselves from photos where their facial features are modified to a preferred size. Another study on the perception of the faces of others confirmed that facial editing can increase perceived attractiveness when features are edited to certain proportions~\cite{przylipiak2018impact}.

An additional self-face perception study looked at the perception of personality, intelligence, emotion and appeal from facial traits in both unmodified faces and faces modified with AR. The participants thought they looked significantly different with filters versus without filters, even when the change to their face was minor. Additionally, facial feature modification changed users' perceptions of the personality traits their modified faces conveyed~\cite{Fribourg_Peillard_McDonnell_2021}. A similar study of traditionally-edited faces assessed their perceived aggressiveness, likeability, sociability, trustworthiness, attractiveness, authenticity, and masculinity or femininity. In this study, participants edited their own photos for attractiveness, and then those photos were evaluated by others. These images were found to be more attractive than the originals, but the judgments of the other traits varied~\cite{parsa2022social}.

These previous works for both traditionally-filtered and AR-filtered faces studied the perceived facial attractiveness or perceived personality traits of the modified faces. In some cases, the impact of these perceptions on self-identity was assessed. Our study takes the perception of filtered faces in a different direction: we study if filtered faces are becoming a ``new normal'' in what people think faces should look like.

\section{METHODS}
We created six Institutional Review Board (IRB)-approved surveys to better understand how people perceive filtered faces. Details of these surveys follow.

\textbf{Image Dataset.} We created a dataset of filtered images and their original, unfiltered counterparts. 485 images of faces of unique subjects were selected from a publicly-available dataset of face images taken at the University of Notre Dame. We chose this dataset because it is one of the standard datasets of controlled facial images used across human biometrics research, including in the NIST evaluations of facial recognition models~\cite{phillips2005overview}~\cite{phillips2007frvt}. All images were taken indoors and the subjects were instructed to have a blank stare expression. The images were captured with a 4 Megapixel Canon PowerShot G2 camera. The dataset included 301 males and 184 females. The race breakdown was 345 White, 80 Asian, 2 Middle Eastern, 12 Hispanic, 8 Black, 14 South Asian and 24 unknown.

\begin{figure}
  \begin{subfigure}[t]{0.48\textwidth}
    \vskip 0pt 
    \includegraphics[width=\textwidth]{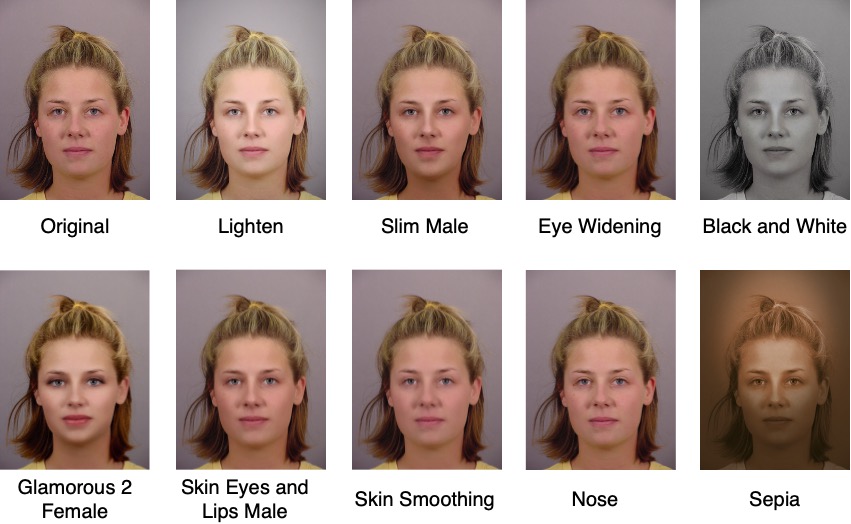}
    \label{fig:portraitprofilters}  
  \end{subfigure}
  \hspace{4pt}
  \begin{subfigure}[t]{0.48\textwidth}
    \vskip 0pt
    \includegraphics[width=\textwidth]{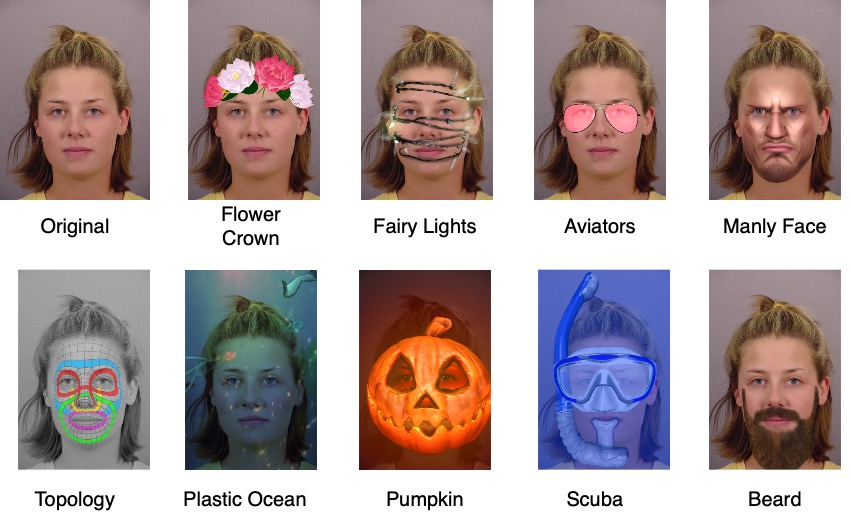}
    \label{fig:deeparfilters}  
  \end{subfigure}
  \vspace{-15pt}
  \caption{The traditional (top) and AR (bottom) filters used in our dataset}
  \Description{In this figure, the same image of a woman is shown with all of the different filters used in the study applied. The first set of filters are the nine traditional edit filters. These include ``lighten'' which lightens the image, ``slim male'' which slims the face, ``eye widening'' which widens the eyes, ``black and white'' which makes the image black and white, ``glamorous 2 female'' which slims the face and applies makeup, ``skin eyes and lips male'' which makes small beautifications to the skin, eyes, and lips, ``skin smoothing'' which smooths the skin, ``nose'' which makes the nose smaller, and ``sepia'' which turns in the image into a sepia image. The second set of filters are the nine augmented reality filters. These include ``flower crown'' which adds a digital flower crown to the forehead, ``fairy lights'' which wraps digital Christmas lights around the face, ``aviators'' which affixes digital pink aviator sunglasses over the eyes, ``manly face'' which makes the face look manly and frightening, ``topology'' which turns the image black and white and puts a face topology graphic over the face, ``plastic ocean'' which makes the image look like the person is underwater with plastic floating around, ``pumpkin'' which puts a jack-o-lantern over the face, ``scuba'' which puts a scuba mask over the person's face and turns the image color tones blue, and ``beard'' which affixes a brown beard to the person's face.}
  \label{fig:allfilters}
 \vspace{-15pt}
\end{figure}

We filtered each of these 485 images with two different sets of facial filters: traditional and augmented reality. We distinguish between these types of facial filtering because of their unique styles: traditional facial filters change the color tones of the image or beautify the face, while augmented reality filters overlay digital objects to the face or add a virtual mask. We acknowledge that in the wild there are many AR filters that are similar to traditional filters, modifying facial structure or applying makeup. Because of this overlap, we chose to focus on AR filters that are more distinct from traditional filters. We also compare these different filtering techniques because while the traditional filters have been around for upwards of thirty years, the augmented reality filters have been around fewer than ten. Distinguishing between the two enables us to consider if the longevity of a filtering technique impacts perception of faces edited in that style. For the traditional filters, we chose to use \href{https://www.anthropics.com/portraitpro/}{PortraitPro Studio Max}, a photo editing software similar to Photoshop that offered convenience due to its ability to batch-apply face filters to large numbers of photos at once. Our AR filters were applied from the DeepAR free filter pack. We applied these to the images using a program we wrote using the \href{https://www.deepar.ai/augmented-reality-sdk}{DeepAR SDK}. All of the filters used are shown in Figure \ref{fig:allfilters}. 

\textbf{Survey Questions.}
We wanted to know if filtered images looked normal or strange to the survey participants (\textbf{RQ1}). However, we also wanted to assess if filters determined to be strange-looking have an inherent strange-ness to them or if the strange-ness stems from unfamiliarity with the filter (\textbf{RQ2}). Additionally, because some of the filters produce realistic-looking images, do people even realize they are looking at a filtered image (\textbf{RQ3})? We explored these questions with the following two-alternative forced choice (2AFC) and Likert scale phrasings:
\begin{itemize}
    \item Are you familiar with this style of image? Response options: \emph{Yes, No}
    \item Does this image look strange? Response options: \emph{Yes, No}
    \item Does the image of this person look digitally altered? Response options: \emph{Yes, No}
    \item On a scale of 1 to 5, how similar is this style of image to what you've seen before? \emph{1: This is completely unlike any style of photo I've seen before. 5: This is very similar to styles of photos I've seen before}
    \item On a scale of 1 to 5, how strange does this image look? \emph{1: Not strange at all. 5: Very strange}
    \item On a scale of 1 to 5, how digitally altered does the image of this person look? \emph{1: Not digitally altered at all. 5: Heavily digitally altered}
\end{itemize}

These surveys are modeled after surveys for studying human facial perception in social psychology. Studies on the other race effect~\cite{o1994structural}, facial attractiveness~\cite{rhodes2003fitting}, and judgements of personality traits from facial appearance~\cite{willis2006first}~\cite{todorov2009evaluating} have used Likert scale surveys to measure human evaluation of faces. Previous studies of human perception of virtual and augmented reality faces have also employed Likert scale surveys to assess perceived traits from the faces~\cite{Ferstl_McDonnell_2018}~\cite{javornik2017mirror}~\cite{Fribourg_Peillard_McDonnell_2021}. This study also draws inspiration from Likert scale surveys in the field of biometrics, including collecting first impressions of faces to train a neural network to predict how crowds assign social attributes to faces~\cite{mccurrie2017predicting} and comparing human and algorithmic abilities to detect facial retouching~\cite{bharati_connors_vatsa_singh_bowyer_2022}.

Each of the six surveys was taken by a different subset of participants so that a participant's experience with one survey wouldn't bias their responses in a future survey. Consequently, we cannot claim causality from the survey results (e.g. we cannot not claim that images filtered with a particular filter are viewed as strange \textit{because} the filter is unfamiliar). However, considering the results of the surveys in tandem can still provide insight into how these images are being perceived. Additionally, the use of six different survey questions provided us with multiple views into how filtered faces are being perceived in comparison to unedited faces.

We performed pilot studies to validate the phrasings of the survey questions, iterating on different phrasings if needed. If for a particular survey the first question phrasing elicited responses with no clear pattern, then we iterated on the question phrasings until we found clear patterns elicited in the responses.

\textbf{Survey Logistics.}
The surveys were hosted on Amazon Mechanical Turk in order to easily recruit a large sample of participants. Each survey asked only one of the above questions. Participants were limited to taking only one of the six surveys. The surveys were restricted to participants in the United States. Each participant was asked to answer that survey's question for a random selection of 100 images from the dataset of original, traditionally-filtered, and AR-filtered images. A fixation cross was shown in between images. The survey took approximately 10 minutes to complete and participants were compensated 1 USD. Enough participants were recruited for each survey such that each image was seen at least 5 times: with the size of our dataset, this meant that at least 461 participants were required per survey. The total number of participants per survey was 517 for \emph{Are you familiar with this style of image?}, 477 for \emph{Does this image look strange?}, 479 for \emph{Does the image of this person look digitally altered?}, 503 for \emph{How similar is this style of image to what you've seen before?}, 464 for \emph{How strange does this image look?}, and 465 for \emph{How digitally altered does the image of this person look?} Users responded to the survey by pressing the `F' key on their keyboard for \emph{yes} and the `J' key for \emph{no} in the case of the 2AFC surveys, and by pressing a number from `1' to `5' on their keyboard for the Likert surveys. Each survey contained 5 control questions interspersed amongst the other survey questions in which the users were asked to hit the letter `A' on their keyboard. The control questions were meant to catch users who repetitively hit keys to submit answers without paying attention to the questions. Participants had to pass 3 of the 5 control questions in order for their survey responses to be accepted. Demographic data regarding age, gender identity, and social media usage was collected.

\textbf{Baseline Surveys.}
We additionally conducted baseline surveys in which the same survey questions were asked but only the original images were shown. The purpose of the baseline surveys was to assess how the original images are perceived without the influence or bias of the filtered images mixed into the dataset. Each baseline survey had between 25 and 38 participants. Results of baseline surveys are included in the appendix.

\section{RESULTS}

\begin{figure*}
\vspace{-10pt}
    \begin{subfigure}[b]{0.32\textwidth}
        \includegraphics[width=\textwidth]{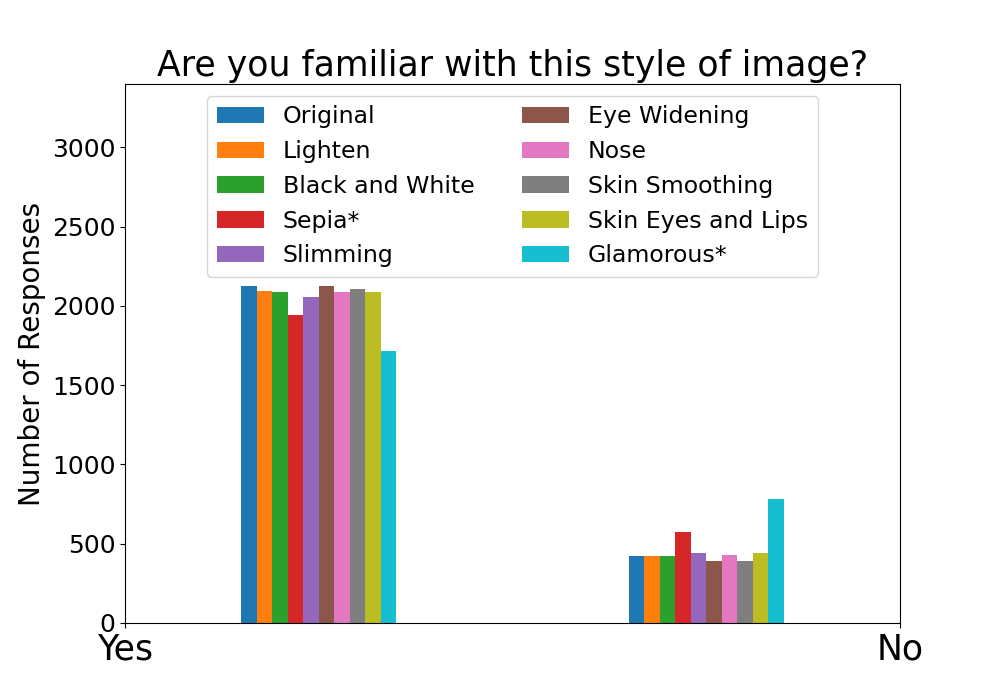}
        \label{fig:normal-2afc-portrait}
    \end{subfigure}
    \begin{subfigure}[b]{0.32\textwidth}
        \includegraphics[width=\textwidth]{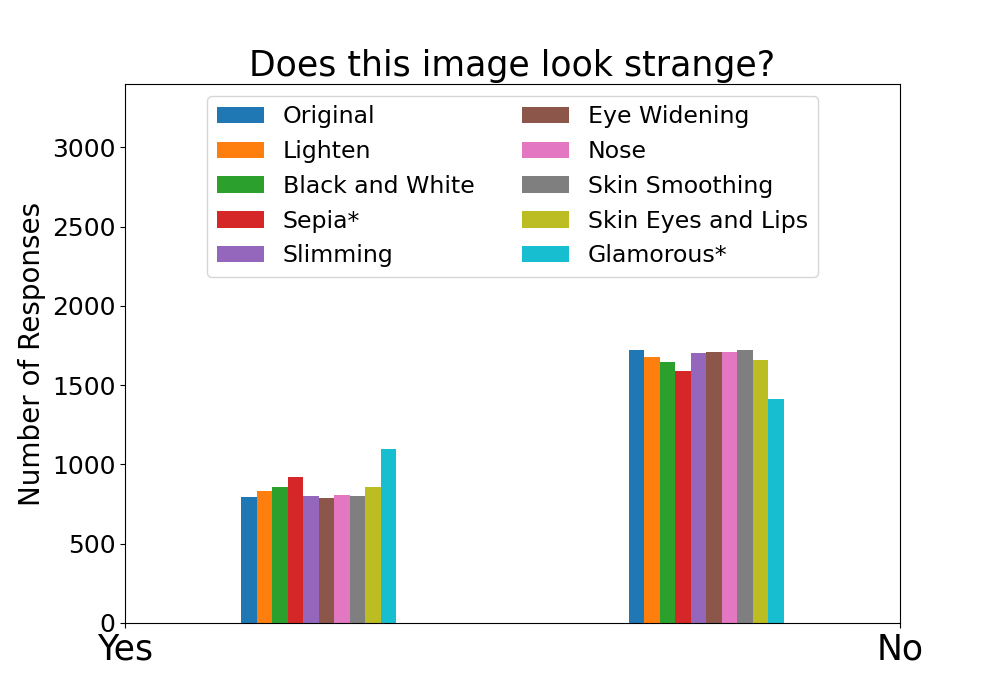}
        \label{fig:unusual-2afc-portrait}
    \end{subfigure}
    \begin{subfigure}[b]{0.32\textwidth}
        \includegraphics[width=\textwidth]{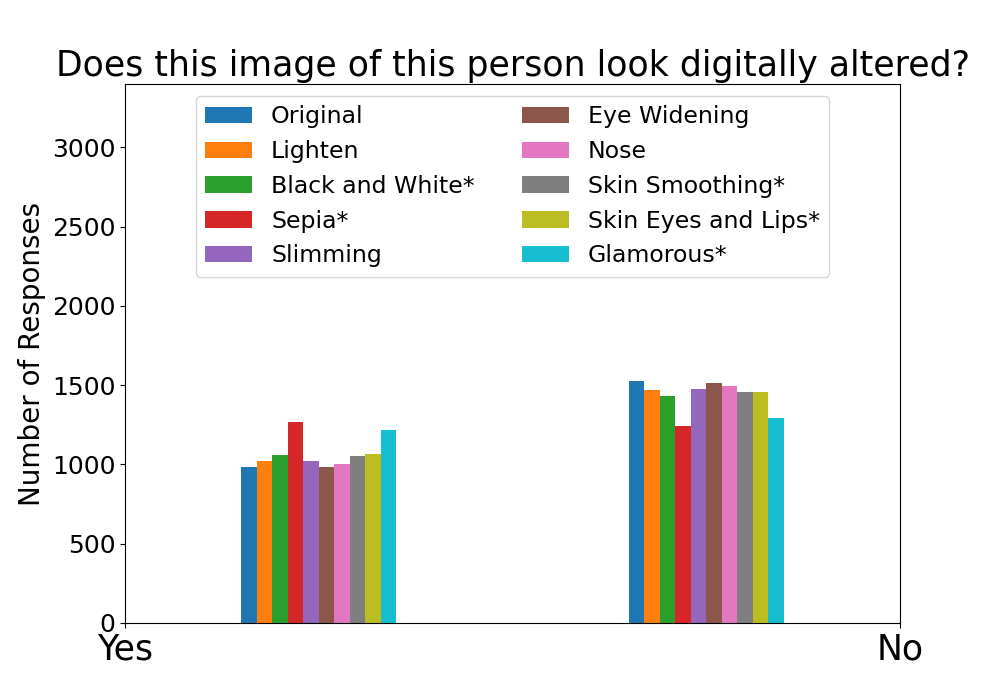}
        \label{fig:edited-2afc-portrait}
    \end{subfigure}
    \vspace{-15pt}
    \caption{Traditional filter vs. original results for the 2AFC surveys. All traditional filters produce similar results to the originals. Filters yielding significantly different results from the originals on a Chi-squared test are indicated with an asterisk (p $<$ 0.5).}
    \Description{This figure shows three bar graphs for the results of the three two-alternative forced choice surveys comparing the traditional filters with the original images. The first is the results of ``Are you familiar with this style of image?'' for all of the traditional edit filters, where all filters and the original images have more yes than no responses. The second is the results for ``Does this image look strange?'' where all filters and the original images have more no responses than yes responses. The third is the results for ``Does this image of this person look digitally altered?'' where there are more no than yes responses for all of the filters and the originals.}
    \label{fig:portrait-2afc-sub-results}
    \vspace{-10pt}
\end{figure*}

\begin{figure*}
    \begin{subfigure}[b]{0.32\textwidth}
        \includegraphics[width=\textwidth]{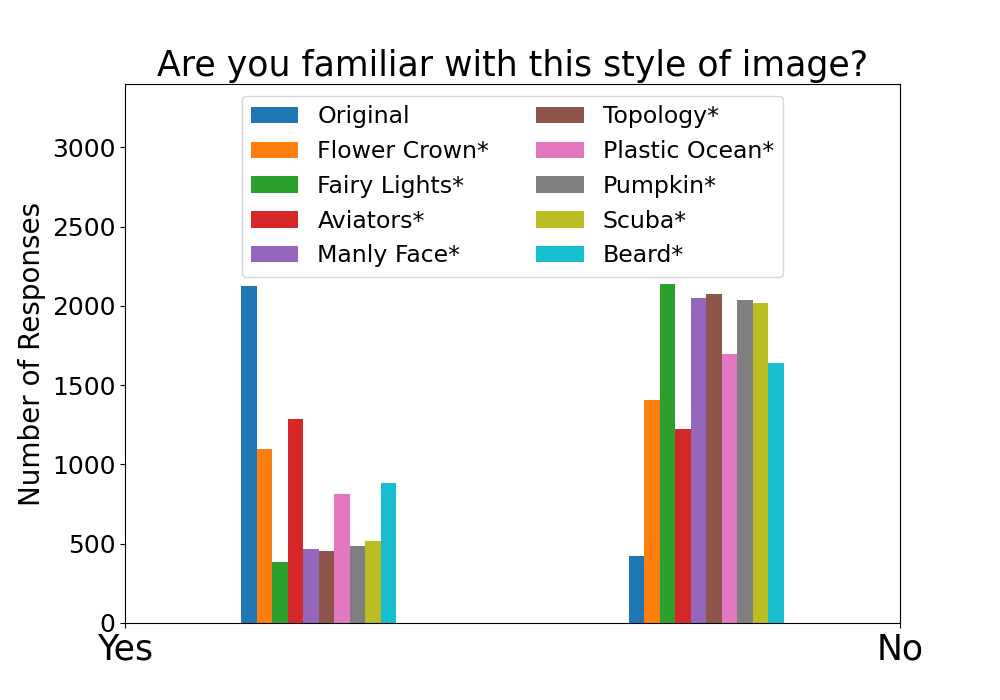}
        \label{fig:normal-2afc-deepar}
    \end{subfigure}
    \begin{subfigure}[b]{0.32\textwidth}
        \includegraphics[width=\textwidth]{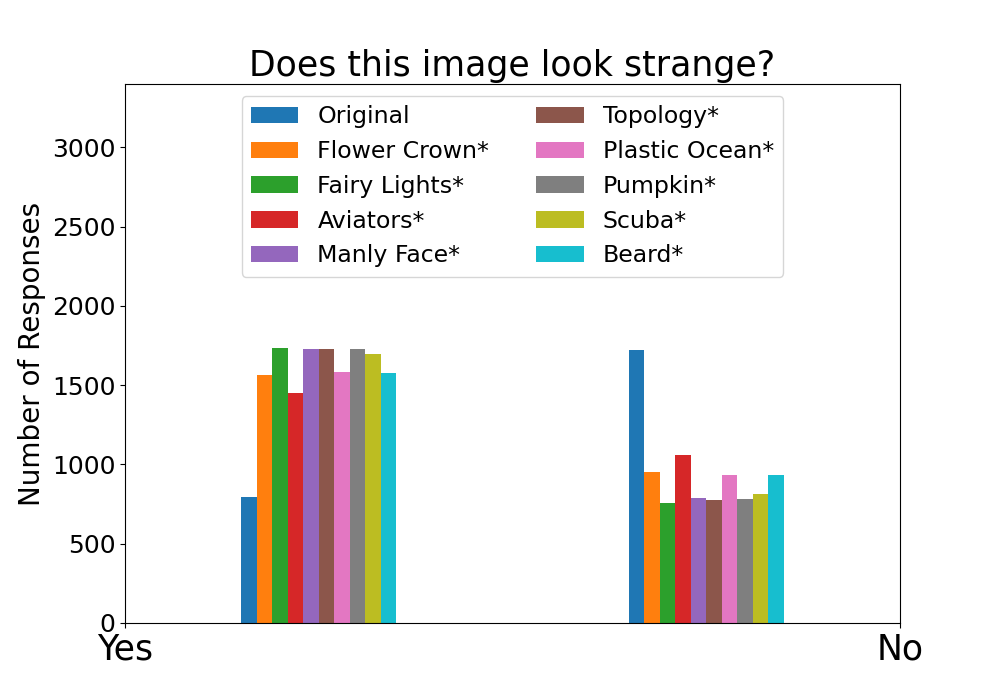}
        \label{fig:unusual-2afc-deepar}
    \end{subfigure}
    \begin{subfigure}[b]{0.32\textwidth}
        \includegraphics[width=\textwidth]{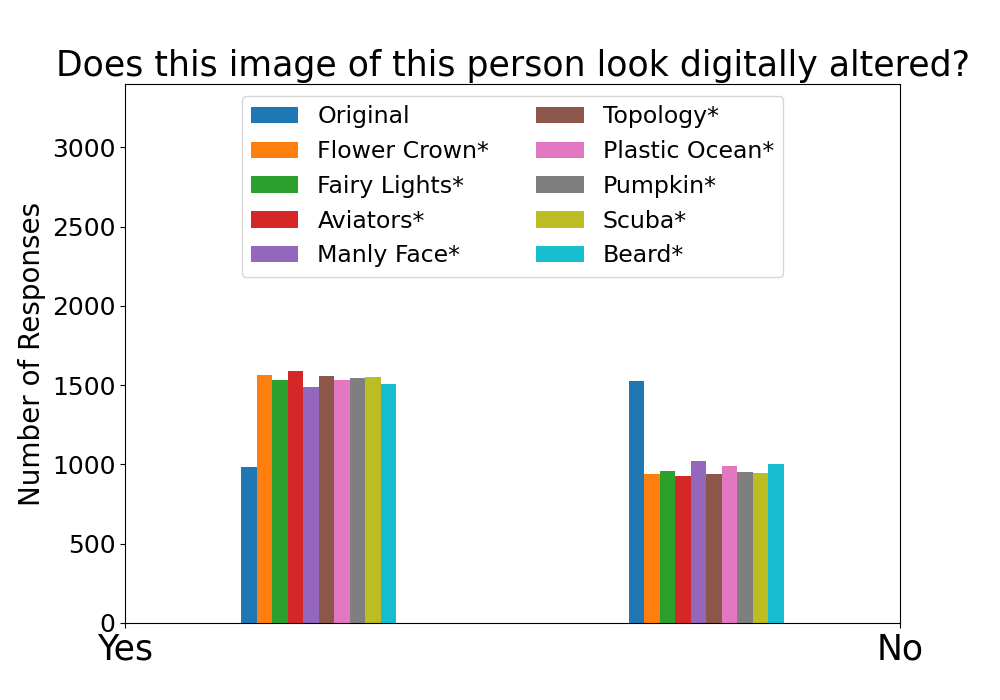}
        \label{fig:edited-2afc-deepar}
    \end{subfigure}
    \vspace{-20pt}
    \caption{AR filter vs. original results for the 2AFC surveys. All AR filters elicited significantly different results on a Chi-squared test (p $<$ 0.5) on all surveys. Significant difference is indicated with an asterisk by the filter name.}
    \Description{This figure shows three bar graphs for the results of the three two-alternative forced choice surveys comparing the AR filters with the original images. The first is the results of ``Are you familiar with this style of image?'' for all of the AR filters, where the original images and aviator filter images have more yes than no responses and all of the other AR filters have the opposite. The second is the results for ``Does this image look strange?'' where the original images have more no responses than yes responses and for all AR filters the results are opposite. The third is the results for ``Does this image of this person look digitally altered?'' where there are more no than yes responses for the originals and all AR filters have more yes than no responses.}
    \label{fig:deepar-2afc-sub-results}
    \vspace{-10pt}
\end{figure*}

\begin{figure*}
    \begin{subfigure}[b]{0.32\textwidth}
        \includegraphics[width=\textwidth]{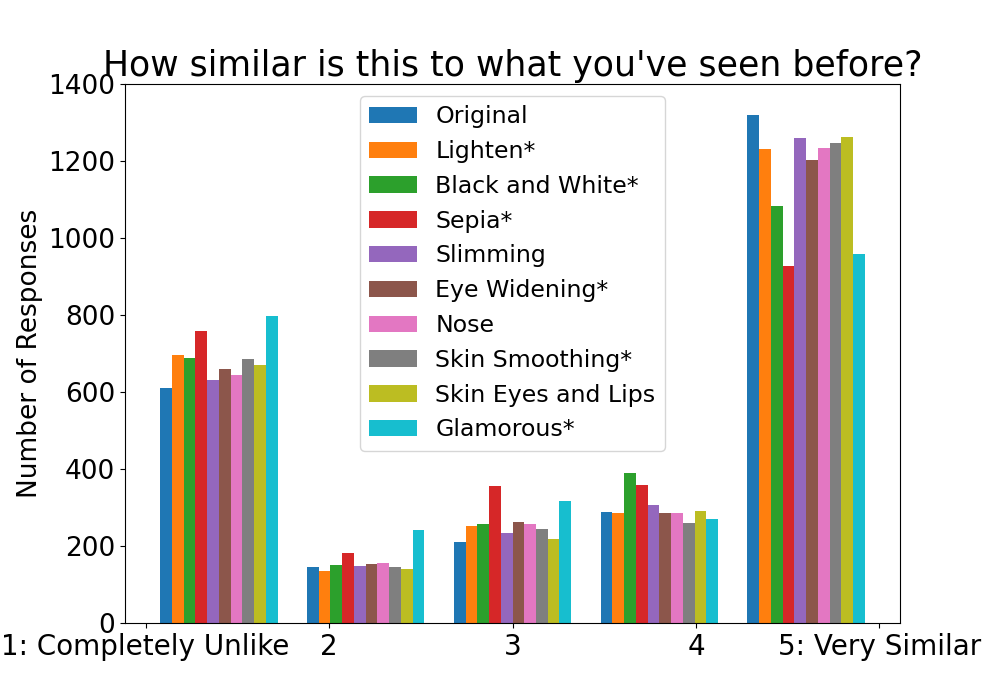}
        \label{fig:normal-likert-portrait}
    \end{subfigure}
    \begin{subfigure}[b]{0.32\textwidth}
        \includegraphics[width=\textwidth]{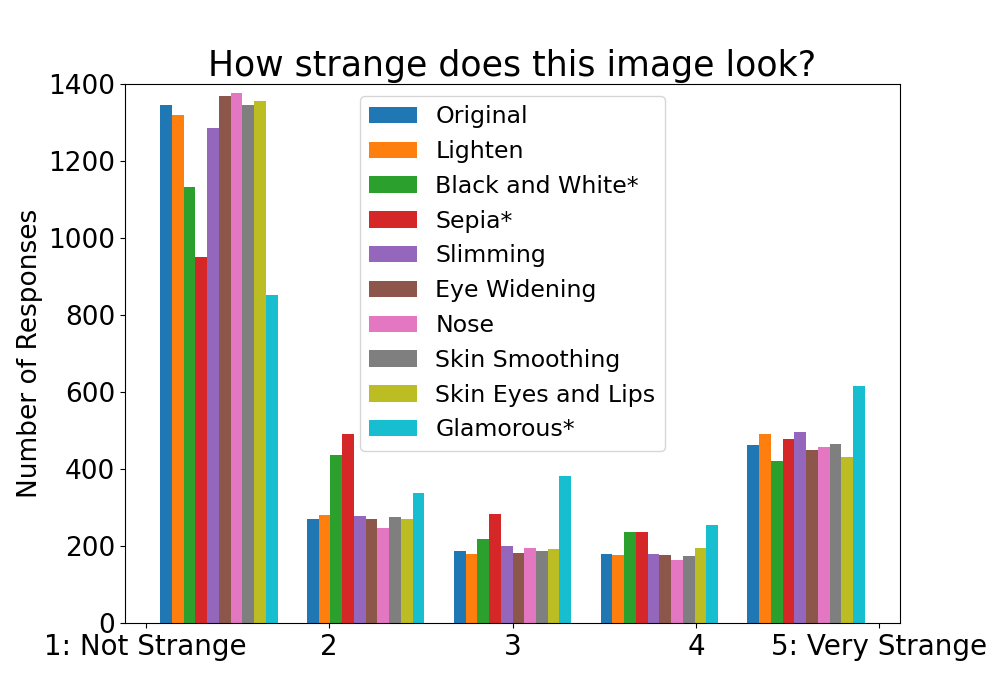}
        \label{fig:unusual-likert-portrait}
    \end{subfigure}
    \begin{subfigure}[b]{0.32\textwidth}
        \includegraphics[width=\textwidth]{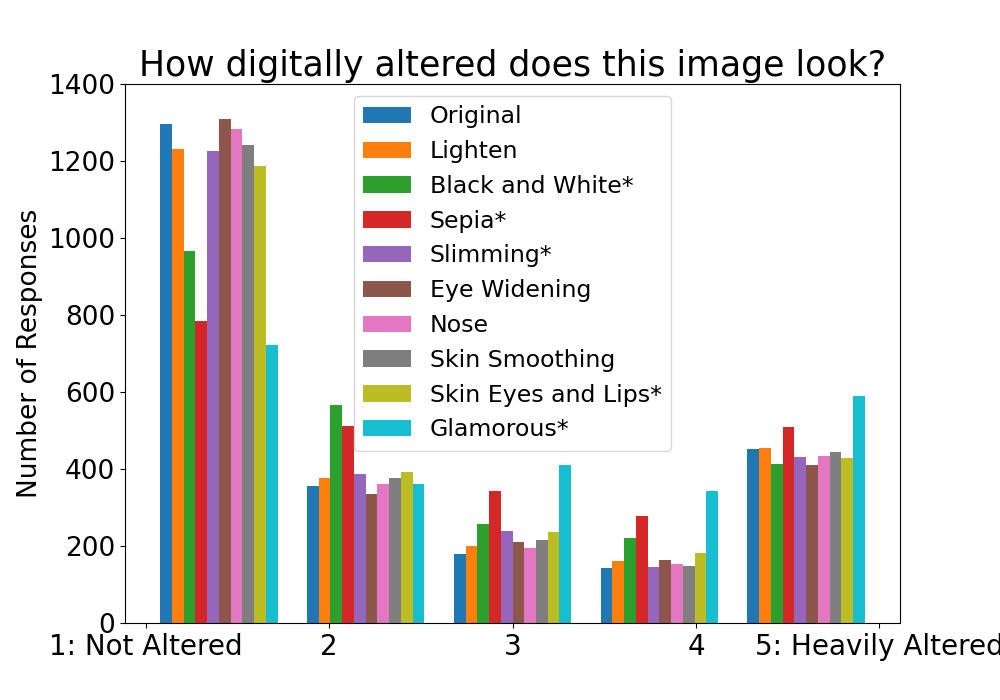}
        \label{fig:edited-likert-portrait}
    \end{subfigure}
    \vspace{-15pt}
    \caption{Traditional filter vs. original results for Likert surveys. Responses to filtered faces vary more from responses to originals on the Likert compared to the 2AFC; significant difference on a Chi-squared test is indicated with an asterisk (p $<$ 0.5).}
    \Description{This figure shows three bar graphs for the results of the three Likert surveys comparing the traditional filters with the original images. The first is the results of ``How similar is this to what you've seen before?'' for all of the traditional edit filters, where all filters and the original images have more 5 (very similar) responses than 1 (completely unlike) responses, and the fewest 2, 3, and 4 responses. The second is the results for ``Does this image look strange?'' where all filters and the original images have more 1 (not strange) responses than 5 (very strange) responses, and the fewest 2, 3, and 4 responses. The third is the results for ``Does this image of this person look digitally altered?'' where all filters and the original images have more 1 (not altered) responses than 5 (heavily altered) responses, and the fewest 2, 3, and 4 responses.}
    \label{fig:portrait-likert-sub-results}
    \vspace{-10pt}
\end{figure*}

\begin{figure*}
    \begin{subfigure}[b]{0.32\textwidth}
        \includegraphics[width=\textwidth]{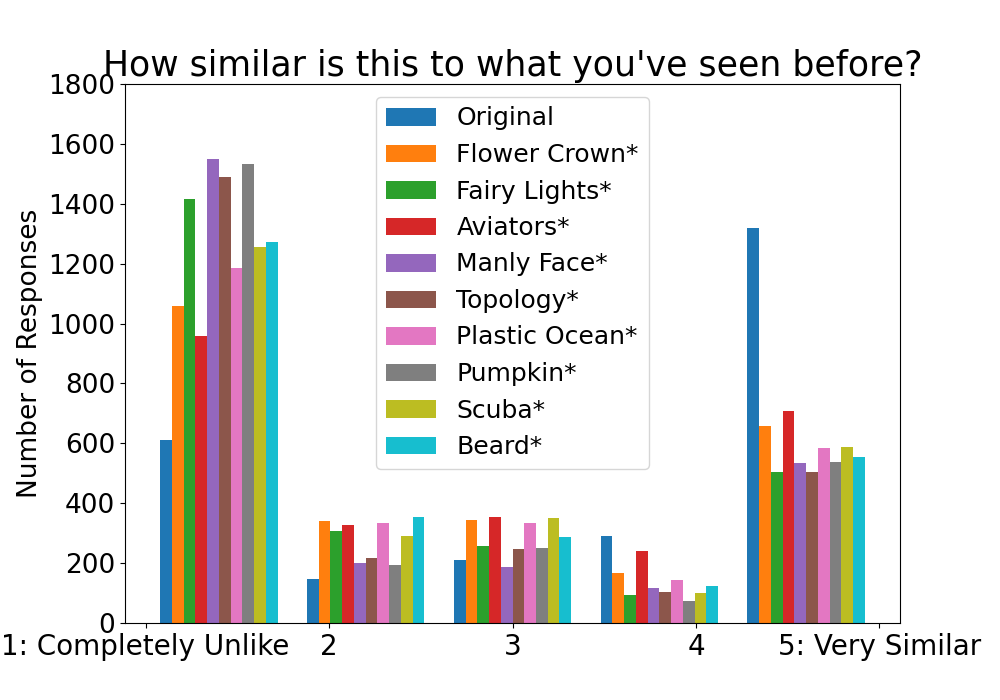}
        \label{fig:normal-likert-deepar}
    \end{subfigure}
    \begin{subfigure}[b]{0.32\textwidth}
        \includegraphics[width=\textwidth]{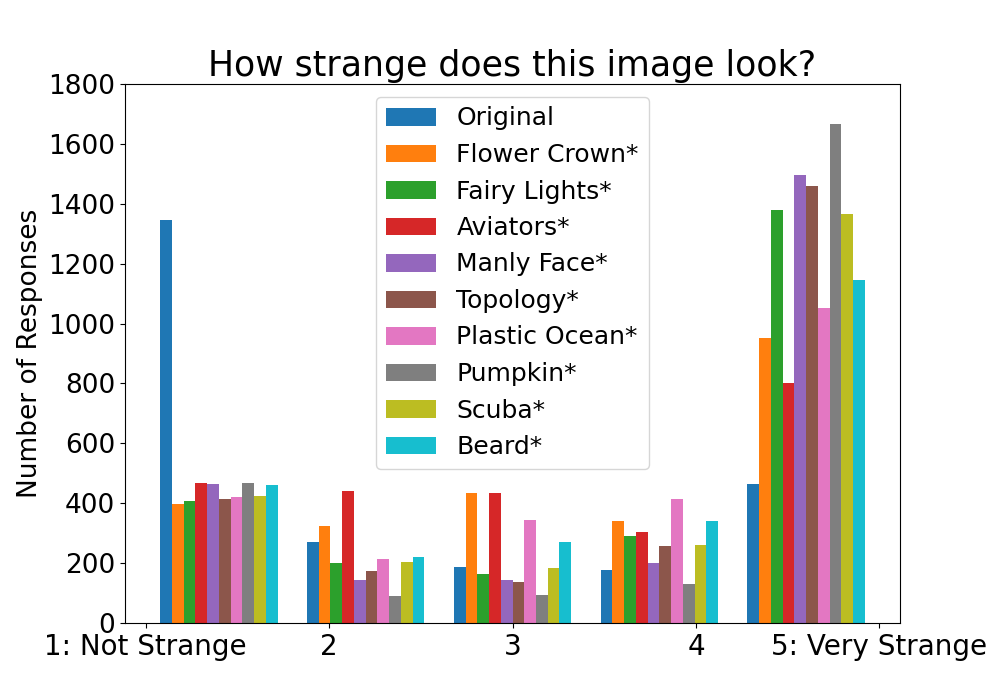}
        \label{fig:unusual-likert-deepar}
    \end{subfigure}
    \begin{subfigure}[b]{0.32\textwidth}
        \includegraphics[width=\textwidth]{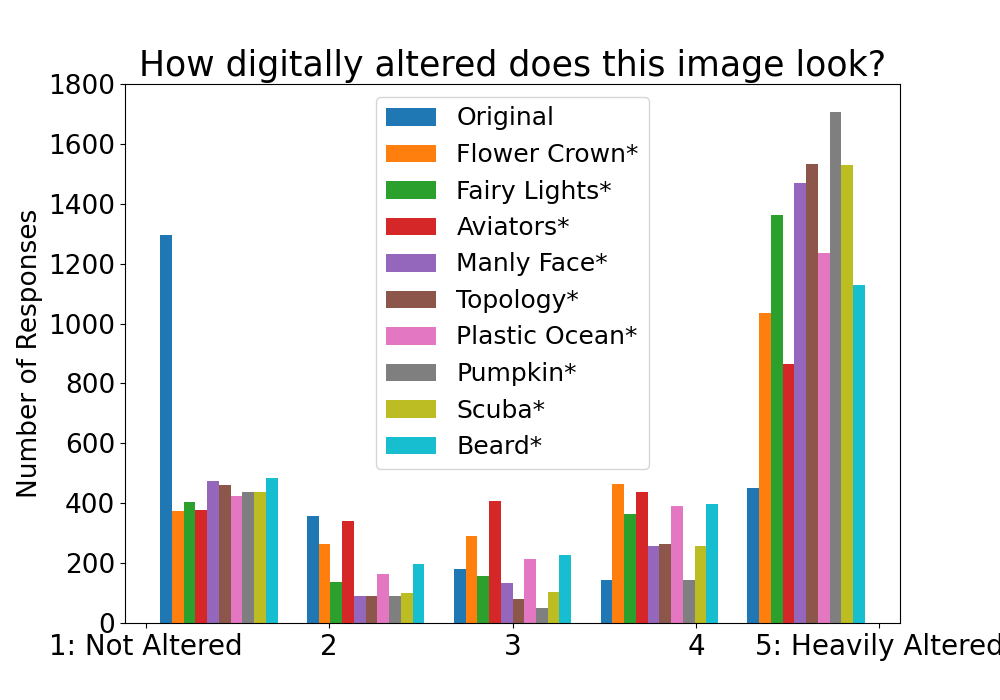}
        \label{fig:edited-likert-deepar}
    \end{subfigure}
     \vspace{-15pt}
    \caption{AR edit vs. original results for Likert surveys. All AR filters elicited significantly different results on a Chi-squared test (p $<$ 0.5) on all surveys. Significant difference is indicated with an asterisk by the filter name.}
     \Description{This figure shows three bar graphs for the results of the three Likert surveys comparing the AR filters with the original images. The first is the results of ``How similar is this to what you've seen before?'' where the original images have more 5 (very similar) responses than 1 (completely unlike) responses, the AR filters have the opposite, and both have few 2, 3, and 4 results. The second is the results for ``Does this image look strange?'' where the original images have more 1 (not strange) responses than 5 (very strange) responses, the AR filters have the opposite, and both have few 2, 3, and 4 responses. The third is the results for ``Does this image of this person look digitally altered?'' where the original images have more 1 (not altered) responses than 5 (heavily altered) responses, the AR filters have the opposite, and both have few 2, 3, and 4 responses.}
    \label{fig:deepar-likert-sub-results}
    \vspace{-8pt}
\end{figure*}

For each survey, we examined the results for the original images, traditionally-edited images, and AR-edited images. Results are visualized by the histograms in Figures \ref{fig:portrait-2afc-sub-results}-\ref{fig:deepar-likert-sub-results}. Additionally, for each survey we performed Chi-squared tests comparing the distributions of responses from the original images and filtered images, for each of the filter types. 

The total recorded amounts of each response option per filter type and per survey, as well as full results of the Chi-squared tests including p-values can be found in the appendix. Although demographic data, including age, gender, and social media usage, was collected, it surprisingly did not provide any meaningful results and is omitted for brevity.

\textit{Traditional Filter Two-Alternative Forced Choice Results}
As with the original images, all traditionally-filtered images were more frequently rated as familiar, not strange, and not digitally altered. However, the Sepia and Glamorous filters had closer results on all surveys compared to the originals, leading to a significant difference (p $<$ 0.05). On the digital alteration survey, the filters Black and White, Skin Smoothing, and Skin Eyes and Lips also elicited significantly different (p $<$ 0.05) responses to the originals due to their closer results.

\textit{Traditional Filter Likert Results}
All of the traditional filters followed similar response distributions to the responses to the originals. On the familiarity survey, both the originals and all traditional filters had mostly ``Very Similar'' (`5') responses followed by ``Completely Unlike'' (`1') responses. In most cases this was followed by `4,' then `3,' then `2' responses. On the strangeness survey, the originals and all traditional filters had mostly ``Not Strange'' (`1') responses followed by ``Very Strange'' (`5') responses. The originals and most traditional filters then had more `2' responses, followed by `3' and `4' responses. More filters were significantly different from the originals compared to the 2AFC surveys, especially Black and White, Sepia, and Glamorous, which were significantly different (p $<$ 0.05) in all three surveys. We acknowledge that significant difference in response distributions is more likely on Likert surveys even across response distributions that appear similar due to the higher number of response options.

\textit{AR Filter Two-Alternative Forced Choice and Likert Results}
On all 2AFC and Likert surveys, every AR filter elicited significantly different results (p $<$ 0.05) from the originals. Moreover, with one exception, on every survey, every AR filter produced an almost inverse distribution of responses from the responses to the originals. In other words, when the originals elicited predominantly `yes' responses on the 2AFC surveys, the AR-filtered faces elicited predominantly `no' responses, and vice versa. On the Likert surveys, the frequency of `5' responses for the original images would resemble the frequency of `1' responses for the AR-filtered images and vice versa. A similar pattern occurred for the `2,' `3,' and `4' responses across the original images and AR-filtered images. The one exception to this is the Aviators filter, which on the familiarity 2AFC survey was more frequently marked as ``familiar'' than ``unfamiliar,'' like the originals. However, the Aviators filter only had slightly more ``familiar'' than ``unfamiliar'' responses, while the originals had many more ``familiar'' than ``unfamiliar'' responses.

\section{DISCUSSION AND CONCLUSION}
The results of our six surveys show that faces filtered with traditional filters are perceived similarly to unedited faces, whereas faces filtered with AR filters are perceived differently from unedited faces. In most cases, the responses to faces filtered with traditional filters are not significantly different from the responses to unedited faces. Even when the responses are significantly different, the distribution of responses to traditionally-filtered faces is relatively similar to the distribution of responses to unedited faces. On the other hand, responses to faces filtered with AR filters are not only significantly different from the responses to unedited faces but have a completely different shaped distribution of responses. In many cases the response distribution for the AR-filtered faces is almost the exact inverse of the response distribution for the original and traditionally edited faces. That these distributions are inverses of each other is notable: people respond to AR-filtered faces opposite from the way that they respond to unedited or traditionally-filtered faces.

Thus it appears that the more traditionally edited faces have become a ``new normal'' in human facial perception, whereas AR-filtered faces have not. This is especially seen in the cases of the Black and White, Skin Eyes and Lips, Slimming, and Skin Smoothing filters which were all identified as digitally altered at somewhat higher (and significantly different) rates than the originals, but on the familiarity and strangeness surveys were not perceived significantly differently from the originals. Because these images are more clearly filtered but still appear normal, our research indicates that these filters in particular are being incorporated into a new normal of what a face should look like. With the knowledge that traditionally edited faces are perceived like unedited faces and AR-filtered faces are not, we recommend future work that further studies the social implications of these findings, including if the perceptual adaptation to filtered faces affects the social judgments we make from facial appearance.

Why are traditionally-filtered faces perceived similarly to unedited faces but AR-filtered faces are not? One possible explanation is the difference in length of time that these filters have existed. Sepia and Black and White images were created with the first cameras and pre-date computer filters. The other traditional filters we selected are in the style of the photo editing that proliferated with Photoshop beginning thirty years ago and was ubiquitous in magazines before the spread of social media. On the other hand, AR filters have only been around for 10 years. Given that the traditional filtering techniques have existed for much longer than the AR filters have, is it just a matter of time until we view AR-filtered images as normal? We recommend future work in continuing to monitor perception of AR-filtered faces over time, to determine if a shift to familiarity will occur.

Another possible explanation for the difference in perception is a difference in the purpose of the filter and the style of its modifications. With the exception of the Black and White, Sepia, and Lighten filters, the traditional filters we used in this study are beautification filters, which serve to enhance the appearance of faces existing in the real world. On the other hand, the AR filters make bigger and more noticeable changes to the image which may have contributed to this difference in perception. Through these more obvious changes, AR filters help to construct alternate realities and stimulate the imagination. Previous research found that creative content creation is a key motivation for AR filter use~\cite{Javornik_Marder_Barhorst_McLean_Rogers_Marshall_Warlop_2022}, and interviews with AR filter creators indicate that many prefer to design ``strange and otherworldly'' filters rather than realistic-looking ones~\cite{pitcher_2020}. With this in mind, our results indicate that these different types of filters are working as they were intended. The faces filtered with the traditional beautification filters still appear normal, allowing them to appear as real-world faces. Conversely, even the most realistic-looking AR filters in our study still caused faces to be perceived as strange, but perhaps this perceived strangeness indicates that the filters are allowing us to escape from reality.

Although both types of filters are working as intended, we must ask: what are the implications of these filters working as intended? Do we want faces filtered with beautification filters to be perceived as normal when those faces don’t actually exist in the real world as they appear in the image? An acne-prone teenager would surely experience a boost of self-confidence with the ability to smooth out their skin before posting a selfie on social media. On the other hand, the Snapchat Dysphoria phenomenon indicates that there is a danger in filtering reality and presenting the results as real. Because the AR-filtered faces in our study were perceived so diametrically differently from unedited faces, it may actually be safer to use these types of filters. They won’t confound our facial perception and cause phenomena like Snapchat Dysphoria in the way that traditional beautification filters would.

One caveat in our support for AR filters is the use of AR beautification filters. We limited the AR filters we used in our study to those that overlay digital objects (Flower Crown, Fairy Lights, Aviators, Plastic Ocean, Pumpkin, Scuba, and Beard) or those that apply unrealistic-looking masks (Manly Face and Topology) to the faces. We did not include AR beautification filters in our study due to their resemblance to the traditional filters we used. However, AR beautification filters take photo editing techniques previously limited to still images and enable them to be applied to a face in real time. We anticipate both the pros and cons of traditional beautification filtering will be amplified when our faces can be filtered in real time. As mentioned in the introduction, this is already possible on Zoom and Alipay, and we encourage future work to study user experiences with real-time beauty filtering on these apps.

Based on our results we recommend AR/VR designers focus more on creating filters that overlay digital objects, transform the face in unrealistic ways, or otherwise promote creativity. We believe that the difference in perception of faces with these filters compared to unfiltered faces indicates that these filters allow us to enter more imaginative worlds. On the other hand, AR beautification filters may continue to further the social concerns already raised by traditional photo beautification techniques. This recommendation is consistent with other literature on AR filter usage. Javornik et. al. found that different motivations for AR filter usage had different effects on well-being: if AR filter usage is driven by a desire to change one’s image to fit an idealized version of oneself that is unattainable offline, then AR filter usage reduces self-acceptance. However, if the AR filter usage aligns with one’s already-existing self image, or helps one actively explore or transform their self-image, it can increase self-acceptance~\cite{Javornik_Marder_Barhorst_McLean_Rogers_Marshall_Warlop_2022}. Additionally, the MIT Technology Review reported that while there are many cases of beautification filters having negative effects on mental health, artistic or creative filters can have a positive effect on mental health~\cite{ryan-mosley_2021}.

One limitation of our study is that our preexisting dataset of faces contained a much higher representation of light skinned faces versus darker complexions. Another limitation is that some of the filters confound with real-world facial modifications, such as applying real makeup or real sunglasses. In fact, we see from our results that of the AR filters, the Aviators filter deviated the most from the results of the other AR filters. Is there a difference in response to faces with real sunglasses or makeup versus faces with virtual sunglasses or makeup? We encourage future work to build on this study by exploring these limitations. We also recommend future work performing the same studies using filters found on popular social media platforms such as Snapchat and Instagram. These specific filters are more ubiquitous than the ones we used in our study, and thus would provide a useful point of comparison. Additionally, we acknowledge the limitation that our study relies on self-reported perceptions which may not reflect actual user behavior. Finally, because our study was limited to participants in the United States, we recommend future work exploring cross-cultural differences in the perception of filtered faces.

\textbf{Conclusion.}
We presented the results of six surveys that measure perceived familiarity, strangeness, and digital alteration of filtered faces. We demonstrated that faces with traditional filters applied are generally perceived similarly to unmodified faces as familiar, not strange, and not digitally altered, but AR-filtered faces are perceived as not familiar, strange, and digitally altered. 
We also must continue to ask: Are AR filters actually more safe to use than traditional filters because they take us out of reality rather than try to perfect it? Or will these AR filters, with time, join the more traditional edits in our perception of what is normal? If so, what are the implications of this?
While we should continue to study whether AR filters will be perceived as ``normal” over time, we recommend for now that AR/VR researchers note that promoting imagination and creativity may be healthier than trying to filter reality.

\begin{acks}
The authors would like to thank Annalisa Szymanski and Daniel Gonzalez Cedre for their feedback on this work.
\end{acks}

\bibliographystyle{ACM-Reference-Format}
\bibliography{sample-base}


\begin{thebibliography}{36}


\ifx \showCODEN    \undefined \def \showCODEN     #1{\unskip}     \fi
\ifx \showDOI      \undefined \def \showDOI       #1{#1}\fi
\ifx \showISBNx    \undefined \def \showISBNx     #1{\unskip}     \fi
\ifx \showISBNxiii \undefined \def \showISBNxiii  #1{\unskip}     \fi
\ifx \showISSN     \undefined \def \showISSN      #1{\unskip}     \fi
\ifx \showLCCN     \undefined \def \showLCCN      #1{\unskip}     \fi
\ifx \shownote     \undefined \def \shownote      #1{#1}          \fi
\ifx \showarticletitle \undefined \def \showarticletitle #1{#1}   \fi
\ifx \showURL      \undefined \def \showURL       {\relax}        \fi
\providecommand\bibfield[2]{#2}
\providecommand\bibinfo[2]{#2}
\providecommand\natexlab[1]{#1}
\providecommand\showeprint[2][]{arXiv:#2}

\bibitem[Ballew and Todorov(2007)]%
        {ballew2007predicting}
\bibfield{author}{\bibinfo{person}{Charles~C Ballew} {and} \bibinfo{person}{Alexander Todorov}.} \bibinfo{year}{2007}\natexlab{}.
\newblock \showarticletitle{Predicting political elections from rapid and unreflective face judgments}.
\newblock \bibinfo{journal}{\emph{Proceedings of the National Academy of Sciences}} \bibinfo{volume}{104}, \bibinfo{number}{46} (\bibinfo{year}{2007}), \bibinfo{pages}{17948--17953}.
\newblock
\urldef\tempurl%
\url{https://doi.org/10.1073/pnas.0705435104}
\showDOI{\tempurl}


\bibitem[Bharati et~al\mbox{.}(2022)]%
        {bharati_connors_vatsa_singh_bowyer_2022}
\bibfield{author}{\bibinfo{person}{Aparna Bharati}, \bibinfo{person}{Emma Connors}, \bibinfo{person}{Mayank Vatsa}, \bibinfo{person}{Richa Singh}, {and} \bibinfo{person}{Kevin Bowyer}.} \bibinfo{year}{2022}\natexlab{}.
\newblock \showarticletitle{In-group and Out-group Performance Bias in Facial Retouching Detection}. In \bibinfo{booktitle}{\emph{2022 IEEE International Joint Conference on Biometrics (IJCB)}}. IEEE, \bibinfo{publisher}{IEEE}, \bibinfo{address}{Abu Dhabi, United Arab Emirates}, \bibinfo{pages}{1--10}.
\newblock
\urldef\tempurl%
\url{https://doi.org/10.1109/ISMAR52148.2021.00064}
\showDOI{\tempurl}


\bibitem[Blair et~al\mbox{.}(2004)]%
        {blair2004influence}
\bibfield{author}{\bibinfo{person}{Irene~V Blair}, \bibinfo{person}{Charles~M Judd}, {and} \bibinfo{person}{Kristine~M Chapleau}.} \bibinfo{year}{2004}\natexlab{}.
\newblock \showarticletitle{The influence of Afrocentric facial features in criminal sentencing}.
\newblock \bibinfo{journal}{\emph{Psychological science}} \bibinfo{volume}{15}, \bibinfo{number}{10} (\bibinfo{year}{2004}), \bibinfo{pages}{674--679}.
\newblock
\urldef\tempurl%
\url{https://doi.org/10.1111/j.0956-7976.2004.00739.x}
\showDOI{\tempurl}


\bibitem[Eberhardt et~al\mbox{.}(2006)]%
        {eberhardt2006looking}
\bibfield{author}{\bibinfo{person}{Jennifer~L Eberhardt}, \bibinfo{person}{Paul~G Davies}, \bibinfo{person}{Valerie~J Purdie-Vaughns}, {and} \bibinfo{person}{Sheri~Lynn Johnson}.} \bibinfo{year}{2006}\natexlab{}.
\newblock \showarticletitle{Looking deathworthy: Perceived stereotypicality of Black defendants predicts capital-sentencing outcomes}.
\newblock \bibinfo{journal}{\emph{Psychological science}} \bibinfo{volume}{17}, \bibinfo{number}{5} (\bibinfo{year}{2006}), \bibinfo{pages}{383--386}.
\newblock
\urldef\tempurl%
\url{https://doi.org/10.1111/j.1467-9280.2006.01716.x}
\showDOI{\tempurl}


\bibitem[Eshiet(2020)]%
        {eshiet2020real}
\bibfield{author}{\bibinfo{person}{Janella Eshiet}.} \bibinfo{year}{2020}\natexlab{}.
\newblock \emph{\bibinfo{title}{“REAL ME VERSUS SOCIAL MEDIA ME:” FILTERS, SNAPCHAT DYSMORPHIA, AND BEAUTY PERCEPTIONS AMONG YOUNG WOMEN}}.
\newblock \bibinfo{thesistype}{Master's\ thesis}. \bibinfo{school}{California State University, San Bernardino}.
\newblock


\bibitem[Fastoso et~al\mbox{.}(2021)]%
        {fastoso2021mirror}
\bibfield{author}{\bibinfo{person}{Fernando Fastoso}, \bibinfo{person}{H{\'e}ctor Gonz{\'a}lez-Jim{\'e}nez}, {and} \bibinfo{person}{Teresa Cometto}.} \bibinfo{year}{2021}\natexlab{}.
\newblock \showarticletitle{Mirror, mirror on my phone: Drivers and consequences of selfie editing}.
\newblock \bibinfo{journal}{\emph{Journal of Business Research}}  \bibinfo{volume}{133} (\bibinfo{year}{2021}), \bibinfo{pages}{365--375}.
\newblock
\urldef\tempurl%
\url{https://doi.org/10.1016/j.jbusres.2021.05.002}
\showDOI{\tempurl}


\bibitem[Felisberti and Musholt(2014)]%
        {felisberti2014self}
\bibfield{author}{\bibinfo{person}{Fatima~M Felisberti} {and} \bibinfo{person}{Kristina Musholt}.} \bibinfo{year}{2014}\natexlab{}.
\newblock \showarticletitle{Self-face perception: Individual differences and discrepancies associated with mental self-face representation, attractiveness and self-esteem}.
\newblock \bibinfo{journal}{\emph{Psychology \& Neuroscience}} \bibinfo{volume}{7}, \bibinfo{number}{2} (\bibinfo{year}{2014}), \bibinfo{pages}{65--72}.
\newblock
\urldef\tempurl%
\url{https://doi.org/10.3922/j.psns.2014.013}
\showDOI{\tempurl}


\bibitem[Ferstl and McDonnell(2018)]%
        {Ferstl_McDonnell_2018}
\bibfield{author}{\bibinfo{person}{Ylva Ferstl} {and} \bibinfo{person}{Rachel McDonnell}.} \bibinfo{year}{2018}\natexlab{}.
\newblock \showarticletitle{A perceptual study on the manipulation of facial features for trait portrayal in virtual agents}. In \bibinfo{booktitle}{\emph{Proceedings of the 18th International Conference on Intelligent Virtual Agents}}. \bibinfo{publisher}{ACM}, \bibinfo{address}{Sydney NSW Australia}, \bibinfo{pages}{281–288}.
\newblock
\showISBNx{978-1-4503-6013-5}
\urldef\tempurl%
\url{https://doi.org/10.1145/3267851.3267891}
\showDOI{\tempurl}


\bibitem[Fribourg et~al\mbox{.}(2021)]%
        {Fribourg_Peillard_McDonnell_2021}
\bibfield{author}{\bibinfo{person}{Rebecca Fribourg}, \bibinfo{person}{Etienne Peillard}, {and} \bibinfo{person}{Rachel McDonnell}.} \bibinfo{year}{2021}\natexlab{}.
\newblock \showarticletitle{Mirror, Mirror on My Phone: Investigating Dimensions of Self-Face Perception Induced by Augmented Reality Filters}. In \bibinfo{booktitle}{\emph{2021 IEEE International Symposium on Mixed and Augmented Reality (ISMAR)}}. \bibinfo{publisher}{IEEE}, \bibinfo{address}{Bari, Italy}, \bibinfo{pages}{470–478}.
\newblock
\showISSN{1554-7868}
\urldef\tempurl%
\url{https://doi.org/10.1109/ISMAR52148.2021.00064}
\showDOI{\tempurl}


\bibitem[Gill(2021)]%
        {gill2021changing}
\bibfield{author}{\bibinfo{person}{Rosalind Gill}.} \bibinfo{year}{2021}\natexlab{}.
\newblock \bibinfo{title}{Changing the perfect picture: Smartphones, social media and appearance pressures}.
\newblock
\newblock
\urldef\tempurl%
\url{https://www.city.ac.uk/\_\_data/assets/pdf\_file/0005/597209/Parliament-Report-web.pdf}
\showURL{%
\tempurl}


\bibitem[Haxby et~al\mbox{.}(2000)]%
        {haxby2000distributed}
\bibfield{author}{\bibinfo{person}{James~V Haxby}, \bibinfo{person}{Elizabeth~A Hoffman}, {and} \bibinfo{person}{M~Ida Gobbini}.} \bibinfo{year}{2000}\natexlab{}.
\newblock \showarticletitle{The distributed human neural system for face perception}.
\newblock \bibinfo{journal}{\emph{Trends in cognitive sciences}} \bibinfo{volume}{4}, \bibinfo{number}{6} (\bibinfo{year}{2000}), \bibinfo{pages}{223--233}.
\newblock
\urldef\tempurl%
\url{https://doi.org/10.1016/s1364-6613(00)01482-0}
\showDOI{\tempurl}


\bibitem[Hunt(2019)]%
        {hunt_2019}
\bibfield{author}{\bibinfo{person}{Elle Hunt}.} \bibinfo{year}{2019}\natexlab{}.
\newblock \bibinfo{title}{Faking it: How selfie dysmorphia is driving people to seek surgery}.
\newblock
\newblock
\urldef\tempurl%
\url{https://www.theguardian.com/lifeandstyle/2019/jan/23/faking-it-how-selfie-dysmorphia-is-driving-people-to-seek-surgery}
\showURL{%
\tempurl}


\bibitem[INDE(2020)]%
        {inde_2020}
\bibfield{author}{\bibinfo{person}{Team INDE}.} \bibinfo{year}{2020}\natexlab{}.
\newblock \bibinfo{title}{The brief history of social media augmented reality filters - inde - the world leader in augmented reality}.
\newblock
\newblock
\urldef\tempurl%
\url{https://www.indestry.com/blog/the-brief-history-of-social-media-ar-filters}
\showURL{%
\tempurl}


\bibitem[Javornik et~al\mbox{.}(2022)]%
        {Javornik_Marder_Barhorst_McLean_Rogers_Marshall_Warlop_2022}
\bibfield{author}{\bibinfo{person}{Ana Javornik}, \bibinfo{person}{Ben Marder}, \bibinfo{person}{Jennifer~Brannon Barhorst}, \bibinfo{person}{Graeme McLean}, \bibinfo{person}{Yvonne Rogers}, \bibinfo{person}{Paul Marshall}, {and} \bibinfo{person}{Luk Warlop}.} \bibinfo{year}{2022}\natexlab{}.
\newblock \showarticletitle{‘What lies behind the filter?’Uncovering the motivations for using augmented reality (AR) face filters on social media and their effect on well-being}.
\newblock \bibinfo{journal}{\emph{Computers in Human Behavior}}  \bibinfo{volume}{128} (\bibinfo{year}{2022}), \bibinfo{pages}{107126}.
\newblock
\showISSN{0747-5632}
\urldef\tempurl%
\url{https://doi.org/10.1016/j.chb.2021.107126}
\showDOI{\tempurl}


\bibitem[Javornik et~al\mbox{.}(2021)]%
        {javornik2021augmented}
\bibfield{author}{\bibinfo{person}{Ana Javornik}, \bibinfo{person}{Ben Marder}, \bibinfo{person}{Marta Pizzetti}, {and} \bibinfo{person}{Luk Warlop}.} \bibinfo{year}{2021}\natexlab{}.
\newblock \showarticletitle{Augmented self-The effects of virtual face augmentation on consumers' self-concept}.
\newblock \bibinfo{journal}{\emph{Journal of Business research}}  \bibinfo{volume}{130} (\bibinfo{year}{2021}), \bibinfo{pages}{170--187}.
\newblock
\urldef\tempurl%
\url{https://doi.org/10.1016/j.jbusres.2021.03.026}
\showDOI{\tempurl}


\bibitem[Javornik and Pizzetti(2017)]%
        {javornik2017mirror}
\bibfield{author}{\bibinfo{person}{Ana Javornik} {and} \bibinfo{person}{Marta Pizzetti}.} \bibinfo{year}{2017}\natexlab{}.
\newblock \showarticletitle{Mirror Mirror on the Wall, Who Is Real of Them All? - the Role of Augmented Self, Expertise and Personalisation in the Experience With Augmented Reality Mirror}.
\newblock \bibinfo{journal}{\emph{ACR North American Advances}} (\bibinfo{year}{2017}).
\newblock
\urldef\tempurl%
\url{https://api.semanticscholar.org/CorpusID:67114806}
\showURL{%
\tempurl}


\bibitem[Jin(2012)]%
        {jin2012virtual}
\bibfield{author}{\bibinfo{person}{Seung-A~Annie Jin}.} \bibinfo{year}{2012}\natexlab{}.
\newblock \showarticletitle{The virtual malleable self and the virtual identity discrepancy model: Investigative frameworks for virtual possible selves and others in avatar-based identity construction and social interaction}.
\newblock \bibinfo{journal}{\emph{Computers in Human Behavior}} \bibinfo{volume}{28}, \bibinfo{number}{6} (\bibinfo{year}{2012}), \bibinfo{pages}{2160--2168}.
\newblock
\urldef\tempurl%
\url{https://doi.org/10.1016/j.chb.2012.06.022}
\showDOI{\tempurl}


\bibitem[Lavrence and Cambre(2020)]%
        {lavrence2020look}
\bibfield{author}{\bibinfo{person}{Christine Lavrence} {and} \bibinfo{person}{Carolina Cambre}.} \bibinfo{year}{2020}\natexlab{}.
\newblock \showarticletitle{“Do I Look Like My Selfie?”: Filters and the Digital-Forensic Gaze}.
\newblock \bibinfo{journal}{\emph{Social Media + Society}} \bibinfo{volume}{6}, \bibinfo{number}{4} (\bibinfo{year}{2020}).
\newblock
\urldef\tempurl%
\url{https://doi.org/10.1177/2056305120955182}
\showDOI{\tempurl}


\bibitem[Little et~al\mbox{.}(2007)]%
        {little2007facial}
\bibfield{author}{\bibinfo{person}{Anthony~C Little}, \bibinfo{person}{Robert~P Burriss}, \bibinfo{person}{Benedict~C Jones}, {and} \bibinfo{person}{S~Craig Roberts}.} \bibinfo{year}{2007}\natexlab{}.
\newblock \showarticletitle{Facial appearance affects voting decisions}.
\newblock \bibinfo{journal}{\emph{Evolution and Human Behavior}} \bibinfo{volume}{28}, \bibinfo{number}{1} (\bibinfo{year}{2007}), \bibinfo{pages}{18--27}.
\newblock
\urldef\tempurl%
\url{https://doi.org/10.1016/j.evolhumbehav.2006.09.002}
\showDOI{\tempurl}


\bibitem[McCurrie et~al\mbox{.}(2017)]%
        {mccurrie2017predicting}
\bibfield{author}{\bibinfo{person}{Mel McCurrie}, \bibinfo{person}{Fernando Beletti}, \bibinfo{person}{Lucas Parzianello}, \bibinfo{person}{Allen Westendorp}, \bibinfo{person}{Samuel Anthony}, {and} \bibinfo{person}{Walter~J Scheirer}.} \bibinfo{year}{2017}\natexlab{}.
\newblock \showarticletitle{Predicting first impressions with deep learning}. In \bibinfo{booktitle}{\emph{2017 12th IEEE International Conference on Automatic Face \& Gesture Recognition (FG 2017)}}. IEEE, \bibinfo{publisher}{IEEE}, \bibinfo{address}{Washington, DC, USA}, \bibinfo{pages}{518--525}.
\newblock
\urldef\tempurl%
\url{https://doi.org/10.1109/FG.2017.147}
\showDOI{\tempurl}


\bibitem[Oosterhof and Todorov(2008)]%
        {oosterhof2008functional}
\bibfield{author}{\bibinfo{person}{Nikolaas~N Oosterhof} {and} \bibinfo{person}{Alexander Todorov}.} \bibinfo{year}{2008}\natexlab{}.
\newblock \showarticletitle{The functional basis of face evaluation}.
\newblock \bibinfo{journal}{\emph{Proceedings of the National Academy of Sciences}} \bibinfo{volume}{105}, \bibinfo{number}{32} (\bibinfo{year}{2008}), \bibinfo{pages}{11087--11092}.
\newblock
\urldef\tempurl%
\url{https://doi.org/10.1073/pnas.0805664105}
\showDOI{\tempurl}


\bibitem[O’Toole et~al\mbox{.}(1994)]%
        {o1994structural}
\bibfield{author}{\bibinfo{person}{Alice~J O’Toole}, \bibinfo{person}{Kenneth~A Deffenbacher}, \bibinfo{person}{Dominique Valentin}, {and} \bibinfo{person}{Herve Abdi}.} \bibinfo{year}{1994}\natexlab{}.
\newblock \showarticletitle{Structural aspects of face recognition and the other-race effect}.
\newblock \bibinfo{journal}{\emph{Memory \& Cognition}}  \bibinfo{volume}{22} (\bibinfo{year}{1994}), \bibinfo{pages}{208--224}.
\newblock
\urldef\tempurl%
\url{https://doi.org/10.3758/BF03208892}
\showDOI{\tempurl}


\bibitem[Parsa et~al\mbox{.}(2022)]%
        {parsa2022social}
\bibfield{author}{\bibinfo{person}{Keon~M Parsa}, \bibinfo{person}{Karina Charipova}, \bibinfo{person}{Eugenia Chu}, {and} \bibinfo{person}{Michael~J Reilly}.} \bibinfo{year}{2022}\natexlab{}.
\newblock \showarticletitle{Social perception of self-enhanced photographs}.
\newblock \bibinfo{journal}{\emph{Facial Plastic Surgery}} \bibinfo{volume}{38}, \bibinfo{number}{03} (\bibinfo{year}{2022}), \bibinfo{pages}{279--284}.
\newblock


\bibitem[Peng(2020)]%
        {peng2020alipay}
\bibfield{author}{\bibinfo{person}{Altman~Yuzhu Peng}.} \bibinfo{year}{2020}\natexlab{}.
\newblock \showarticletitle{Alipay adds “beauty filters” to face-scan payments: a form of patriarchal control over women’s bodies}.
\newblock \bibinfo{journal}{\emph{Feminist Media Studies}} \bibinfo{volume}{20}, \bibinfo{number}{4} (\bibinfo{year}{2020}), \bibinfo{pages}{582--585}.
\newblock
\urldef\tempurl%
\url{https://doi.org/10.1080/14680777.2020.1750779}
\showDOI{\tempurl}


\bibitem[Phillips et~al\mbox{.}(2005)]%
        {phillips2005overview}
\bibfield{author}{\bibinfo{person}{P~Jonathon Phillips}, \bibinfo{person}{Patrick~J Flynn}, \bibinfo{person}{Todd Scruggs}, \bibinfo{person}{Kevin~W Bowyer}, \bibinfo{person}{Jin Chang}, \bibinfo{person}{Kevin Hoffman}, \bibinfo{person}{Joe Marques}, \bibinfo{person}{Jaesik Min}, {and} \bibinfo{person}{William Worek}.} \bibinfo{year}{2005}\natexlab{}.
\newblock \showarticletitle{Overview of the face recognition grand challenge}. In \bibinfo{booktitle}{\emph{2005 IEEE computer society conference on computer vision and pattern recognition (CVPR'05)}}, Vol.~\bibinfo{volume}{1}. IEEE, \bibinfo{publisher}{IEEE}, \bibinfo{address}{San Diego, CA, USA}, \bibinfo{pages}{947--954}.
\newblock
\urldef\tempurl%
\url{https://doi.org/10.1109/CVPR.2005.268}
\showDOI{\tempurl}


\bibitem[Phillips et~al\mbox{.}(2007)]%
        {phillips2007frvt}
\bibfield{author}{\bibinfo{person}{P~Jonathon Phillips}, \bibinfo{person}{W~Todd Scruggs}, \bibinfo{person}{Alice~J O’Toole}, \bibinfo{person}{Patrick~J Flynn}, \bibinfo{person}{Kevin~W Bowyer}, \bibinfo{person}{Cathy~L Schott}, {and} \bibinfo{person}{Matthew Sharpe}.} \bibinfo{year}{2007}\natexlab{}.
\newblock \showarticletitle{FRVT 2006 and ICE 2006 large-scale results}.
\newblock \bibinfo{journal}{\emph{National Institute of Standards and Technology, NISTIR}} \bibinfo{volume}{7408}, \bibinfo{number}{1} (\bibinfo{year}{2007}), \bibinfo{pages}{1}.
\newblock
\urldef\tempurl%
\url{https://doi.org/10.1109/TPAMI.2009.59}
\showDOI{\tempurl}


\bibitem[Pitcher(2020)]%
        {pitcher_2020}
\bibfield{author}{\bibinfo{person}{Laura Pitcher}.} \bibinfo{year}{2020}\natexlab{}.
\newblock \bibinfo{title}{Instagram Filters are Changing the Way We Think About Makeup}.
\newblock
\newblock
\urldef\tempurl%
\url{https://www.teenvogue.com/story/instagram-filters-makeup}
\showURL{%
\tempurl}


\bibitem[Przylipiak et~al\mbox{.}(2018)]%
        {przylipiak2018impact}
\bibfield{author}{\bibinfo{person}{Mateusz Przylipiak}, \bibinfo{person}{Jerzy Przylipiak}, \bibinfo{person}{Robert Terlikowski}, \bibinfo{person}{Emilia Lubowicka}, \bibinfo{person}{Lech Chrostek}, {and} \bibinfo{person}{Andrzej Przylipiak}.} \bibinfo{year}{2018}\natexlab{}.
\newblock \showarticletitle{Impact of face proportions on face attractiveness}.
\newblock \bibinfo{journal}{\emph{Journal of cosmetic dermatology}} \bibinfo{volume}{17}, \bibinfo{number}{6} (\bibinfo{year}{2018}), \bibinfo{pages}{954--959}.
\newblock
\urldef\tempurl%
\url{https://doi.org/10.1111/jocd.12783}
\showDOI{\tempurl}


\bibitem[Rhodes et~al\mbox{.}(2003)]%
        {rhodes2003fitting}
\bibfield{author}{\bibinfo{person}{Gillian Rhodes}, \bibinfo{person}{Linda Jeffery}, \bibinfo{person}{Tamara~L Watson}, \bibinfo{person}{Colin~WG Clifford}, {and} \bibinfo{person}{Ken Nakayama}.} \bibinfo{year}{2003}\natexlab{}.
\newblock \showarticletitle{Fitting the mind to the world: Face adaptation and attractiveness aftereffects}.
\newblock \bibinfo{journal}{\emph{Psychological science}} \bibinfo{volume}{14}, \bibinfo{number}{6} (\bibinfo{year}{2003}), \bibinfo{pages}{558--566}.
\newblock
\urldef\tempurl%
\url{https://doi.org/10.1046/j.0956-7976.2003.psci_1465.x}
\showDOI{\tempurl}


\bibitem[Ryan-Mosley(2021)]%
        {ryan-mosley_2021}
\bibfield{author}{\bibinfo{person}{Tate Ryan-Mosley}.} \bibinfo{year}{2021}\natexlab{}.
\newblock \bibinfo{title}{Beauty filters are changing the way young girls see themselves}.
\newblock
\newblock
\urldef\tempurl%
\url{https://www.technologyreview.com/2021/04/02/1021635/beauty-filters-young-girls-augmented-reality-social-media/}
\showURL{%
\tempurl}


\bibitem[Shein(2021)]%
        {Shein_2021}
\bibfield{author}{\bibinfo{person}{Esther Shein}.} \bibinfo{year}{2021}\natexlab{}.
\newblock \showarticletitle{Filtering for beauty}.
\newblock \bibinfo{journal}{\emph{Commun. ACM}} \bibinfo{volume}{64}, \bibinfo{number}{11} (\bibinfo{date}{Nov} \bibinfo{year}{2021}), \bibinfo{pages}{17–19}.
\newblock
\showISSN{0001-0782, 1557-7317}
\urldef\tempurl%
\url{https://doi.org/10.1145/3484997}
\showDOI{\tempurl}


\bibitem[Todorov et~al\mbox{.}(2005)]%
        {todorov2005inferences}
\bibfield{author}{\bibinfo{person}{Alexander Todorov}, \bibinfo{person}{Anesu~N Mandisodza}, \bibinfo{person}{Amir Goren}, {and} \bibinfo{person}{Crystal~C Hall}.} \bibinfo{year}{2005}\natexlab{}.
\newblock \showarticletitle{Inferences of competence from faces predict election outcomes}.
\newblock \bibinfo{journal}{\emph{Science}} \bibinfo{volume}{308}, \bibinfo{number}{5728} (\bibinfo{year}{2005}), \bibinfo{pages}{1623--1626}.
\newblock
\urldef\tempurl%
\url{https://doi.org/10.1126/science.1110589}
\showDOI{\tempurl}


\bibitem[Todorov et~al\mbox{.}(2009)]%
        {todorov2009evaluating}
\bibfield{author}{\bibinfo{person}{Alexander Todorov}, \bibinfo{person}{Manish Pakrashi}, {and} \bibinfo{person}{Nikolaas~N Oosterhof}.} \bibinfo{year}{2009}\natexlab{}.
\newblock \showarticletitle{Evaluating faces on trustworthiness after minimal time exposure}.
\newblock \bibinfo{journal}{\emph{Social cognition}} \bibinfo{volume}{27}, \bibinfo{number}{6} (\bibinfo{year}{2009}), \bibinfo{pages}{813--833}.
\newblock
\urldef\tempurl%
\url{https://doi.org/10.1521/soco.2009.27.6.813}
\showDOI{\tempurl}


\bibitem[Willis and Todorov(2006)]%
        {willis2006first}
\bibfield{author}{\bibinfo{person}{Janine Willis} {and} \bibinfo{person}{Alexander Todorov}.} \bibinfo{year}{2006}\natexlab{}.
\newblock \showarticletitle{First impressions: Making up your mind after a 100-ms exposure to a face}.
\newblock \bibinfo{journal}{\emph{Psychological science}} \bibinfo{volume}{17}, \bibinfo{number}{7} (\bibinfo{year}{2006}), \bibinfo{pages}{592--598}.
\newblock
\urldef\tempurl%
\url{https://doi.org/10.1111/j.1467-9280.2006.01750.x}
\showDOI{\tempurl}


\bibitem[Wilson(2009)]%
        {Wilson_2009}
\bibfield{author}{\bibinfo{person}{Eric Wilson}.} \bibinfo{year}{2009}\natexlab{}.
\newblock \showarticletitle{Smile and Say ‘No Photoshop’}.
\newblock \bibinfo{journal}{\emph{The New York Times}} (\bibinfo{date}{May} \bibinfo{year}{2009}).
\newblock
\urldef\tempurl%
\url{https://www.nytimes.com/2009/05/28/fashion/28RETOUCH.html}
\showURL{%
\tempurl}


\bibitem[Yim and Park(2019)]%
        {Yim_Park_2019}
\bibfield{author}{\bibinfo{person}{Mark Yi-Cheon Yim} {and} \bibinfo{person}{Sun-Young Park}.} \bibinfo{year}{2019}\natexlab{}.
\newblock \showarticletitle{“I am not satisfied with my body, so I like augmented reality (AR)”: Consumer responses to AR-based product presentations}.
\newblock \bibinfo{journal}{\emph{Journal of Business Research}}  \bibinfo{volume}{100} (\bibinfo{date}{Jul} \bibinfo{year}{2019}), \bibinfo{pages}{581–589}.
\newblock
\showISSN{0148-2963}
\urldef\tempurl%
\url{https://doi.org/10.1016/j.jbusres.2018.10.041}
\showDOI{\tempurl}


\end{thebibliography}

\appendix

\section{Baseline Survey Results}

``Are you familiar with this style of image?,'' ``Does this image look strange?,'' and ``How strange does this image look?'' had 25 survey participants, ``Does the image of this person look digitally altered?'' and ``How digitally altered does this image look?'' had 26 participants, and ``How similar is this to what you've seen before?'' had 38 participants. 

\textit{Are you familiar with this style of image? Results.}
In the baseline study, the original images had around 72.5\% `yes' responses and 27.5\% `no' responses. The original images in the context of the full study had around 83.4\% `yes' responses and 16.6\% `no' responses. These results are significantly different from each other (p $<$ 0.05) despite that on both surveys many more respondents responded `yes' than `no.' With the filtered images mixed in to the dataset, the original images are rated as more familiar. We conjecture that when there are no filtered images mixed in, subjects rate fewer of the unfiltered images as familiar because the question primes them to think there are unfamiliar images in the set. However, the converse could be true: that the presence of less familiar filtered images caused subjects to find the unfiltered images to be more familiar.

\textit{Does this image look strange? Results.}
In the baseline study, the original images had 30\% `yes' responses and 70\% `no' responses. In the full study, the original images had 31.6\% `yes' responses and 68.4\% `no' responses. These two sets of responses were not found to be significantly different from each other on a Chi-Squared test.

\textit{Does this image look digitally altered? Results.}
In the baseline study, the original images had around 35.8\% `yes' and 64.2\% `no' responses. In the context of the full study, the original images had 39.2\% `yes' responses and 60.8\% `no' responses. This is significantly different (p $<$ 0.05) from the responses to the original images in the context of the full survey. Although the results were significantly different, they were similar. However, the presence of clearly filtered images in the dataset caused respondents to think that slightly more unedited images were edited.

\textit{How familiar does this style of image look? Results.}
In the baseline study, the original images had around 17\% `1', 14\% `2', 16\% `3', 19\% `4', and 33\% `5'. In the full study, the original responses followed a distribution of around 23-26\% `1', around 5-6\% `2', 8-10\% `3', around 11\% `4', and around 45-50\% `5'. The baseline survey produced responses significantly different from the originals (p $<$ 0.05) on a Chi-Squared test. In this case, the main difference in responses came from more of the `2', `3', and `4', responses being selected in the baseline study. We speculate that the presence of filtered images in the dataset caused respondents to select the response options at the extremes, because the different filters caused the images to all look very different from each other. In the baseline survey where only unfiltered images were included, because the images were all more relatively similar to each other, that could have encouraged the participants to use more of the response options in the middle.

\textit{How strange does this image look? Results.}
The original image responses in the main study were 55\%  `1', 11\% `2', 8\% `3', 7\% `4', and 19\% `5'. In the baseline study, the original images had around 44\% `1', 11\% `2', 11\% `3', 13\% `4', and 22\% `5' responses. The distribution is significantly different from the response distribution for the original images in the main study. We observe two main differences between the response distributions. First, the baseline study had both more `2,' `3,' and `4,' responses compared to the responses to the originals in the full study, and the responses skewed more towards looking more strange. Second, when there were no filtered images in the dataset, users responded that the original images looked more strange. This is similar to what we saw in the similarity surveys, where participants selected more of the `2', `3', and `4' options on the baseline survey than on the full survey, and participants answered ``Yes'' or `5' for more unfiltered images on the baseline study than in the full study. We propose the same explanation for these differences as we did with the similarity survey.

\textit{How digitally altered does this image look? Results.}
In the main study, the original images had 45-53\% `1' responses, 14-15\% `2' responses, 6-11\% `3', and `4' responses, and 18\% `5' responses. In the baseline study, the original images had around 39\% `1', 23\% `2', 18\% `3', 9\% `4', and 11\% `5' responses. This response distribution was significantly different (p $<$ 0.05) from the distribution of responses for the original images in the full study. The main difference is more `2,' `3,' and `4,' responses, especially `2' responses, in the baseline study. As mentioned earlier, we see this trend in the other two baseline surveys, and we speculate that when filtered images are present, the more extreme differences in the appearance of the images encouraged participants to choose the more extreme options on the Likert scale.

\section{Full Survey Results}
In this section we provide the number of each type of response to each survey, broken down by filter.

\subsection{Are you familiar with this style of image? Results}

Original:
Yes: $n = 2125$, No: $n = 422$\\
PortraitPro:
Yes: $n = 18312$, No: $n = 4293$\\
DeepAR:
Yes: $n = 6385$, No: $n = 16288$\\
Flower Crown:
Yes: $n = 1097$, No: $n = 1406$\\
Fairy Lights:
Yes: $n = 386$, No: $n = 2140$\\
Aviators:
Yes: $n = 1284$, No: $n = 1223$\\
Manly Face:
Yes: $n = 465$, No: $n = 2048$\\
Topology:
Yes: $n = 453$, No: $n = 2078$\\
Plastic Ocean:
Yes: $n = 815$, No: $n = 1696$\\
Pumpkin:
Yes: $n = 484$, No: $n = 2039$\\
Scuba:
Yes: $n = 520$, No: $n = 2017$\\
Beard:
Yes: $n = 881$, No: $n = 1641$\\
Lighten:
Yes: $n = 2094$, No: $n = 423$\\
Slim Male:
Yes: $n = 2058$, No: $n = 442$\\
Eye Widening:
Yes: $n = 2127$, No: $n = 391$\\
Black and White:
Yes: $n = 2086$, No: $n = 421$\\
Glamorous:
Yes: $n = 1716$, No: $n = 782$\\
Skin eyes and lips:
Yes: $n = 2089$, No: $n = 440$\\
Skin Smoothing:
Yes: $n = 2107$, No: $n = 393$\\
Nose:
Yes: $n = 2090$, No: $n = 426$\\
Sepia:
Yes: $n = 1945$, No: $n = 575$\\
Original images on the baseline survey:
Yes: $n = 1812$, No: $n = 688$\\

\subsection{Does this image look strange? Results}
Original:
Yes: $n = 797$, No: $n = 1725$\\
PortraitPro:
Yes: $n = 7761$, No: $n = 14832$\\
DeepAR:
Yes: $n = 14795$, No: $n = 7795$\\
Flower Crown:
Yes: $n = 1563$, No: $n = 953$\\
Fairy Lights:
Yes: $n = 1734$, No: $n = 760$\\
Aviators:
Yes: $n = 1452$, No: $n = 1059$\\
Manly Face:
Yes: $n = 1729$, No: $n = 786$\\
Topology:
Yes: $n = 1728$, No: $n = 774$\\
Plastic Ocean:
Yes: $n = 1583$, No: $n = 934$\\
Pumpkin:
Yes: $n = 1730$, No: $n = 782$\\
Scuba:
Yes: $n = 1696$, No: $n = 815$\\
Beard:
Yes: $n = 1580$, No: $n = 932$\\
Lighten:
Yes: $n = 831$, No: $n = 1677$\\
Slim Male:
Yes: $n = 802$, No: $n = 1706$\\
Eye Widening:
Yes: $n = 790$, No: $n = 1709$\\
Black and White:
Yes: $n = 856$, No: $n = 1649$\\
Glamorous:
Yes: $n = 1095$, No: $n = 1412$\\
Skin eyes and lips:
Yes: $n = 858$, No: $n = 1657$\\
Skin Smoothing:
Yes: $n = 798$, No: $n = 1725$\\
Nose:
Yes: $n = 809$, No: $n = 1710$\\
Sepia:
Yes: $n = 922$, No: $n = 1587$\\
Original images on the baseline survey:
Yes: $n = 751$, No: $n = 1749$\\

\subsection{Does this image look digitally altered? Results}
Original:
Yes: $n = 982$, No: $n = 1525$\\
PortraitPro:
Yes: $n = 9701$, No: $n = 12850$\\
DeepAR:
Yes: $n = 13872$, No: $n = 8670$\\
Flower Crown:
Yes: $n = 1563$, No: $n = 937$\\
Fairy Lights:
Yes: $n = 1532$, No: $n = 957$\\
Aviators:
Yes: $n = 1589$, No: $n = 925$\\
Manly Face:
Yes: $n = 1489$, No: $n = 1020$\\
Topology:
Yes: $n = 1557$, No: $n = 939$\\
Plastic Ocean:
Yes: $n = 1530$, No: $n = 991$\\
Pumpkin:
Yes: $n = 1548$, No: $n = 954$\\
Scuba:
Yes: $n = 1554$, No: $n = 946$\\
Beard:
Yes: $n = 1510$, No: $n = 1001$\\
Lighten:
Yes: $n = 1024$, No: $n = 1471$\\
Slim Male:
Yes: $n = 1024$, No: $n = 1479$\\
Eye Widening:
Yes: $n = 985$, No: $n = 1515$\\
Black and White:
Yes: $n = 1062$, No: $n = 1433$\\
Glamorous:
Yes: $n = 1217$, No: $n = 1295$\\
Skin eyes and lips:
Yes: $n = 1063$, No: $n = 1458$\\
Skin Smoothing:
Yes: $n = 1053$, No: $n = 1459$\\
Nose:
Yes: $n = 1004$, No: $n = 1497$\\
Sepia:
Yes: $n = 1269$, No: $n = 1243$\\
Original images on the baseline survey:
Yes: $n = 896$, No: $n = 1604$\\

\subsection{How familiar does this style of image look? Results}
Responses are in a Likert scale from ``1: completely unfamiliar'' to ``5: very similar.''

Original:
1: $n = 610$, 2: $n = 146$, 3: $n = 211$, 4: $n = 289$, 5: $n = 1320$\\
PortraitPro:
1: $n = 6231$, 2: $n = 1453$, 3: $n = 2397$, 4: $n = 2731$, 5: $n = 10406$\\
DeepAR:
1: $n = 11713$, 2: $n = 2564$, 3: $n = 2613$, 4: $n = 1162$, 5: $n = 5175$\\
Flower Crown:
1: $n = 1058$, 2: $n = 341$, 3: $n = 344$, 4: $n = 167$, 5: $n = 659$\\
Fairy Lights:
1: $n = 1416$, 2: $n = 308$, 3: $n = 257$, 4: $n = 94$, 5: $n = 503$\\
Aviators:
1: $n = 957$, 2: $n = 326$, 3: $n = 354$, 4: $n = 241$, 5: $n = 708$\\
Manly Face:
1: $n = 1550$, 2: $n = 199$, 3: $n = 187$, 4: $n = 116$, 5: $n = 535$\\
Topology:
1: $n = 1488$, 2: $n = 218$, 3: $n = 248$, 4: $n = 104$, 5: $n = 503$\\
Plastic Ocean:
1: $n = 1184$, 2: $n = 334$, 3: $n = 333$, 4: $n = 143$, 5: $n = 583$\\
Pumpkin:
1: $n = 1532$, 2: $n = 194$, 3: $n = 252$, 4: $n = 72$, 5: $n = 539$\\
Scuba:
1: $n = 1256$, 2: $n = 291$, 3: $n = 351$, 4: $n = 100$, 5: $n = 589$\\
Beard:
1: $n = 1272$, 2: $n = 353$, 3: $n = 287$, 4: $n = 125$, 5: $n = 556$\\
Lighten:
1: $n = 697$, 2: $n = 134$, 3: $n = 251$, 4: $n = 286$, 5: $n = 1230$\\
Slim Male:
1: $n = 630$, 2: $n = 149$, 3: $n = 235$, 4: $n = 307$, 5: $n = 1260$\\
Eye Widening:
1: $n = 660$, 2: $n = 154$, 3: $n = 263$, 4: $n = 286$, 5: $n = 1203$\\
Black and White:
1: $n = 689$, 2: $n = 151$, 3: $n = 257$, 4: $n = 389$, 5: $n = 1084$\\
Glamorous:
1: $n = 797$, 2: $n = 241$, 3: $n = 317$, 4: $n = 269$, 5: $n = 958$\\
Skin eyes and lips:
1: $n = 669$, 2: $n = 139$, 3: $n = 217$, 4: $n = 290$, 5: $n = 1263$\\
Skin Smoothing:
1: $n = 686$, 2: $n = 146$, 3: $n = 244$, 4: $n = 261$, 5: $n = 1247$\\
Nose:
1: $n = 645$, 2: $n = 156$, 3: $n = 256$, 4: $n = 285$, 5: $n = 1235$\\
Sepia:
1: $n = 758$, 2: $n = 183$, 3: $n = 357$, 4: $n = 358$, 5: $n = 926$\\
Original images on the baseline survey:
1: $n = 675$, 2: $n = 588$, 3: $n = 663$, 4: $n = 749$, 5: $n = 1325$\\

\subsection{How strange does this image look? Results}
Responses are in a Likert scale from ``1: not strange at all'' to ``5: very strange.''
Original:
1: $n = 1345$, 2: $n = 269$, 3: $n = 186$, 4: $n = 178$, 5: $n = 463$\\
PortraitPro:
1: $n = 10991$, 2: $n = 2886$, 3: $n = 2017$, 4: $n = 1792$, 5: $n = 4310$\\
DeepAR:
1: $n = 3921$, 2: $n = 2009$, 3: $n = 2195$, 4: $n = 2532$, 5: $n = 11316$\\
Flower Crown:
1: $n = 396$, 2: $n = 323$, 3: $n = 433$, 4: $n = 340$, 5: $n = 952$\\
Fairy Lights:
1: $n = 407$, 2: $n = 202$, 3: $n = 162$, 4: $n = 289$, 5: $n = 1379$\\
Aviators:
1: $n = 466$, 2: $n = 441$, 3: $n = 433$, 4: $n = 304$, 5: $n = 800$\\
Manly Face:
1: $n = 465$, 2: $n = 142$, 3: $n = 143$, 4: $n = 199$, 5: $n = 1497$\\
Topology:
1: $n = 415$, 2: $n = 174$, 3: $n = 136$, 4: $n = 258$, 5: $n = 1460$\\
Plastic Ocean:
1: $n = 420$, 2: $n = 215$, 3: $n = 343$, 4: $n = 413$, 5: $n = 1052$\\
Pumpkin:
1: $n = 467$, 2: $n = 89$, 3: $n = 92$, 4: $n = 129$, 5: $n = 1665$\\
Scuba:
1: $n = 423$, 2: $n = 203$, 3: $n = 183$, 4: $n = 259$, 5: $n = 1367$\\
Beard:
1: $n = 462$, 2: $n = 220$, 3: $n = 270$, 4: $n = 341$, 5: $n = 1144$\\
Lighten:
1: $n = 1320$, 2: $n = 281$, 3: $n = 179$, 4: $n = 177$, 5: $n = 491$\\
Slim Male:
1: $n = 1287$, 2: $n = 277$, 3: $n = 199$, 4: $n = 180$, 5: $n = 497$\\
Eye Widening:
1: $n = 1370$, 2: $n = 271$, 3: $n = 181$, 4: $n = 176$, 5: $n = 450$\\
Black and White:
1: $n = 1133$, 2: $n = 436$, 3: $n = 218$, 4: $n = 236$, 5: $n = 422$\\
Glamorous:
1: $n = 853$, 2: $n = 337$, 3: $n = 382$, 4: $n = 255$, 5: $n = 616$\\
Skin eyes and lips:
1: $n = 1355$, 2: $n = 270$, 3: $n = 192$, 4: $n = 194$, 5: $n = 432$\\
Skin Smoothing:
1: $n = 1346$, 2: $n = 276$, 3: $n = 187$, 4: $n = 173$, 5: $n = 465$\\
Nose:
1: $n = 1377$, 2: $n = 246$, 3: $n = 195$, 4: $n = 164$, 5: $n = 458$\\
Sepia:
1: $n = 950$, 2: $n = 492$, 3: $n = 284$, 4: $n = 237$, 5: $n = 479$\\
Original images on the baseline survey:
1: $n = 1092$, 2: $n = 269$, 3: $n = 278$, 4: $n = 318$, 5: $n = 543$\\

\subsection{How digitally altered does this image look? Results}
Responses are in a Likert scale from ``1: not digitally altered'' to ``5: heavily digitally altered.''
Original:
1: $n = 1297$, 2: $n = 357$, 3: $n = 179$, 4: $n = 144$, 5: $n = 451$\\
PortraitPro:
1: $n = 9952$, 2: $n = 3666$, 3: $n = 2305$, 4: $n = 1796$, 5: $n = 4118$\\
DeepAR:
1: $n = 3871$, 2: $n = 1468$, 3: $n = 1662$, 4: $n = 2971$, 5: $n = 11867$\\
Flower Crown:
1: $n = 373$, 2: $n = 265$, 3: $n = 290$, 4: $n = 464$, 5: $n = 1034$\\
Fairy Lights:
1: $n = 405$, 2: $n = 138$, 3: $n = 156$, 4: $n = 363$, 5: $n = 1364$\\
Aviators:
1: $n = 377$, 2: $n = 339$, 3: $n = 408$, 4: $n = 438$, 5: $n = 864$\\
Manly Face:
1: $n = 474$, 2: $n = 91$, 3: $n = 133$, 4: $n = 258$, 5: $n = 1471$\\
Topology:
1: $n = 460$, 2: $n = 89$, 3: $n = 79$, 4: $n = 263$, 5: $n = 1534$\\
Plastic Ocean:
1: $n = 423$, 2: $n = 162$, 3: $n = 214$, 4: $n = 390$, 5: $n = 1237$\\
Pumpkin:
1: $n = 439$, 2: $n = 89$, 3: $n = 50$, 4: $n = 142$, 5: $n = 1705$\\
Scuba:
1: $n = 437$, 2: $n = 99$, 3: $n = 105$, 4: $n = 256$, 5: $n = 1530$\\
Beard:
1: $n = 483$, 2: $n = 196$, 3: $n = 227$, 4: $n = 397$, 5: $n = 1128$\\
Lighten:
1: $n = 1232$, 2: $n = 376$, 3: $n = 201$, 4: $n = 162$, 5: $n = 455$\\
Slim Male:
1: $n = 1226$, 2: $n = 387$, 3: $n = 238$, 4: $n = 146$, 5: $n = 431$\\
Eye Widening:
1: $n = 1309$, 2: $n = 334$, 3: $n = 210$, 4: $n = 163$, 5: $n = 411$\\
Black and White:
1: $n = 967$, 2: $n = 567$, 3: $n = 257$, 4: $n = 221$, 5: $n = 414$\\
Glamorous:
1: $n = 722$, 2: $n = 360$, 3: $n = 410$, 4: $n = 344$, 5: $n = 590$\\
Skin eyes and lips:
1: $n = 1187$, 2: $n = 393$, 3: $n = 236$, 4: $n = 182$, 5: $n = 429$\\
Skin Smoothing:
1: $n = 1241$, 2: $n = 377$, 3: $n = 216$, 4: $n = 148$, 5: $n = 444$\\
Nose:
1: $n = 1284$, 2: $n = 360$, 3: $n = 194$, 4: $n = 152$, 5: $n = 435$\\
Sepia:
1: $n = 784$, 2: $n = 512$, 3: $n = 343$, 4: $n = 278$, 5: $n = 509$\\
Original images on the baseline survey:
1: $n = 1001$, 2: $n = 607$, 3: $n = 477$, 4: $n = 228$, 5: $n = 287$\\

\section{Full Chi-Squared Test Results}

In this section we provide the results of the Chi-squared tests for each survey, comparing the results by filter to the results for the original images.

\begin{table}[H]
  \caption{Are you familiar with this style of image? Statistical test results}
  \label{tab:normal2afc}
  \begin{tabular}{ccl}
    \toprule
    Filter&Chi-Square Value&p-value\\
    \midrule
    Flower Crown & 855.64 & \num{4.3e-188}\\
    Fairy Lights & 2353.62 & 0.0\\
    Aviators & 595.76 & \num{1.40e-131}\\
    Manly Face & 2131.60 & 0.0\\
    Topology & 2178.68 & 0.0\\
    Plastic Ocean & 1347.75 & \num{4.75e-295}\\
    Pumpkin & 2091.96 & 0.0\\
    Scuba & 2014.45 & 0.0\\
    Beard & 1233.01 & \num{4.10e-270}\\
    Lighten & 0.04 & 0.85\\
    Slimming & 1.02 & 0.31\\
    Eye Widening & 0.94 & 0.33\\
    Black and White & 0.03 & 0.86\\
    Glamorous & 149.92 & \num{1.80e-34}\\
    Skin Eyes and Lips & 0.56 & 0.45\\
    Skin Smoothing & 0.61 & 0.43\\
    Nose & 0.10 & 0.76\\
    Sepia & 30.90 & 2.71\\
    Original images on baseline survey & 87.56 & \num{8.17e-21}\\
  \bottomrule
\end{tabular}
\end{table}

\begin{table}[H]
  \caption{Does this look strange? Statistical test results}
  \label{tab:unusual2afc}
  \begin{tabular}{ccl}
    \toprule
    Filter&Chi-Square Value&p-value\\
    \midrule
    Flower Crown & 469.94 & \num{3.30e-104}\\
    Fairy Lights & 719.97 & \num{1.36e-158}\\
    Aviators & 349.00 & \num{6.99e-78} \\
    Manly Face & 693.52 & \num{7.66e-153}\\
    Topology & 703.61 & \num{4.90e-155}\\
    Plastic Ocean & 493.64 & \num{2.32e-109}\\
    Pumpkin & 697.67 & \num{9.58e-154}\\
    Scuba & 648.75 & \num{4.16e-143}\\
    Beard & 493.33 & \num{2.69e-109} \\
    Lighten & 1.28 & 0.26\\
    Slimming & 0.07 & 0.80\\
    Eye Widening & 0.0 & 1.0\\
    Black and White & 3.64 & 0.06\\
    Glamorous & 77.61 & \num{1.26e-18}\\
    Skin Eyes and Lips & 3.50 & 0.06\\
    Skin Smoothing & 0.0 & 1.0\\
    Nose & 0.13 & 0.72\\
    Sepia & 14.59 & 0.00\\
    Original images on baseline survey & 1.36 & 0.24\\
  \bottomrule
\end{tabular}
\end{table}

\begin{table}[H]
  \caption{Does this look digitally altered? Statistical test results}
  \label{tab:edited2afc}
  \begin{tabular}{ccl}
    \toprule
    Filter&Chi-Square Value&p-value\\
    \midrule
    Flower Crown & 272.13 & \num{3.90e-61}\\
    Fairy Lights & 249.36 & \num{3.59e-56}\\
    Aviators & 289.28 & \num{7.14e-65}\\
    Manly Face & 203.43 & \num{3.74e-46}\\
    Topology & 268.63 & \num{2.25e-60}\\
    Plastic Ocean & 231.99 & \num{2.20e-52}\\
    Pumpkin & 257.23 & \num{6.88e-58}\\
    Scuba & 263.76 & \num{2.60e-59}\\
    Beard & 219.73 & \num{1.04e-49}\\
    Lighten & 1.75 & 0.19\\
    Slimming & 1.51 & 0.22\\
    Eye Widening & 0.02 & 0.89\\
    Black and White & 5.82 & 0.02\\
    Glamorous & 43.49 & \num{4.26e-11}\\
    Skin, eyes, and lips & 4.55 & 0.03\\
    Skin smoothing & 3.82 & 0.05\\
    Nose & 0.46 & 0.50\\
    Sepia & 64.86 & \num{8.05e-16}\\
    Original images on baseline survey & 5.78 & 0.02\\
  \bottomrule
\end{tabular}
\end{table}

\begin{table}[H]
  \caption{How familiar are you with this style of image? Statistical test results}
  \label{tab:normallikert}
  \begin{tabular}{ccl}
    \toprule
    Filter&Chi-Square Value&p-value\\
    \midrule
    Flower Crown & 483.69 & 2.26\\
    Fairy Lights & 848.41 & \num{2.51e-182}\\
    Aviators & 370.69 & \num{5.96e-79}\\
    Manly Face & 824.74 & \num{3.36e-177} \\
    Topology & 837.86 & \num{4.83e-180}\\
    Plastic Ocean & 619.42 & \num{9.71e-133} \\
    Pumpkin & 865.80 & \num{4.28e-186}\\
    Scuba & 678.35 & \num{1.69e-145} \\
    Beard & 706.39 & \num{1.44e-151} \\
    Lighten & 12.87 & 0.01\\
    Slimming & 3.58 & 0.47\\
    Eye Widening & 13.31 & 0.01\\
    Black and White & 47.32 & \num{1.31e-09} \\
    Glamorous & 127.69 & \num{1.21e-26}\\
    Skin Eyes and Lips & 4.24 & 0.37\\
    Skin Smoothing & 10.34 & 0.04\\
    Nose & 8.50 & 0.07\\
    Sepia & 134.17 & \num{5.00e-28}\\
    Original images on baseline survey & 418.33 & \num{3.05e-89}\\
  \bottomrule
\end{tabular}
\end{table}

\begin{table}[H]
  \caption{How strange does this look? Statistical test results}
  \label{tab:unusuallikert}
  \begin{tabular}{ccl}
    \toprule
    Filter&Chi-Square Value&p-value\\
    \midrule
    Flower Crown & 840.43 &  \num{1.34e-180}\\
    Fairy Lights & 995.28 &  \num{3.77e-214}\\
    Aviators & 689.72 &  \num{5.86e-148}\\
    Manly Face & 1019.36 &  \num{2.27e-219}\\
    Topology & 1051.14 &  \num{2.95e-226}\\
    Plastic Ocean & 859.83 &  \num{8.42e-185}\\
    Pumpkin & 1234.49 &  \num{5.31e-266}\\
    Scuba & 951.64 &  \num{1.08e-204}\\
    Beard & 791.64 &  \num{4.96e-170}\\
    Lighten &  1.45 &  0.84\\
    Slimming &  3.05 &  0.55\\
    Eye Widening &  0.49 &  0.97\\
    Black and White & 70.25 &  \num{2.00e-14}\\
    Glamorous & 220.78 &  \num{1.27e-46}\\
    Skin, eyes, and lips &  1.90 &  0.76\\
    Skin smoothing &  0.16 &  1.00\\
    Nose &  2.22 &  0.70\\
    Sepia & 162.43 & \num{4.41e-34}\\
    Original images on baseline survey & 89.69 & \num{1.53e-18}\\
  \bottomrule
\end{tabular}
\end{table}

\begin{table}[H]
  \caption{How digitally altered does this look? Statistical test results}
  \label{tab:editedlikert}
  \begin{tabular}{ccl}
    \toprule
    Filter&Chi-Square Value&p-value\\
    \midrule
    Flower Crown & 948.42 &  \num{5.36e-204}\\
    Fairy Lights & 1119.82 &  \num{3.83e-241}\\
    Aviators & 873.64 &  \num{8.55e-188}\\
    Manly Face & 1120.81 &  \num{2.33e-241}\\
    Topology & 1224.20 &  \num{9.04e-264}\\
    Plastic Ocean & 999.81 & \num{3.91e-215}\\
    Pumpkin & 1387.15 &  \num{4.23e-299}\\
    Scuba & 1210.85 & \num{7.09e-261}\\
    Beard & 833.37 & \num{4.53e-179}\\
    Lighten &  4.51 &  0.34\\
    Slimming & 12.02 & 0.02\\
    Eye Widening & 6.32 &  0.18\\
    Black and White & 127.61 &  \num{1.26e-26}\\
    Glamorous & 354.89 & \num{1.54e-75}\\
    Skin, eyes, and lips & 19.41 &  0.00065\\
    Skin smoothing &  5.36 &  0.25\\
    Nose &  1.18 &  0.88\\
    Sepia & 251.69 &  \num{2.82e-53}\\
    Original images on baseline survey & 288.20 & \num{3.80e-61}\\
  \bottomrule
\end{tabular}
\end{table}

\end{document}